%% file: main.tex
\PassOptionsToPackage{numbers,compress}{natbib}
\documentclass{article}
\usepackage[preprint]{neurips_2026}
\usepackage[utf8]{inputenc}
\usepackage[T1]{fontenc}
\usepackage{hyperref}
\usepackage{url}
\usepackage{booktabs}
\usepackage{amsfonts}
\usepackage{amssymb}
\usepackage{amsmath}
\usepackage{amsthm}
\usepackage{microtype}
\usepackage{xcolor}
\usepackage{graphicx}
\usepackage{algorithm}
\usepackage{algorithmic}
\usepackage{multirow}
\usepackage[shortlabels]{enumitem}
\usepackage{tabularx}
\usepackage{float}
\usepackage{caption}

\setlist{noitemsep,topsep=2pt,partopsep=0pt,parsep=0pt}
\usepackage{titlesec}
\titlespacing*{\section}{0pt}{9pt plus 2pt minus 1pt}{4pt plus 1pt}
\titlespacing*{\subsection}{0pt}{7pt plus 2pt minus 1pt}{3pt plus 1pt}
\theoremstyle{definition}
\newtheorem{definition}{Definition}

\makeatletter
\def\thm@space@setup{%
  \thm@preskip=4pt plus 1pt minus 1pt
  \thm@postskip=2pt plus 1pt minus 1pt
}
\makeatother

\newcommand{\AskBack}{Asking-Back}
\newcommand{\llmjudge}{LLM-as-judge}

\newlength{\figheight}

\newcommand{\NumStudents}{63}
\newcommand{\NumFamilies}{3}
\newcommand{\NumConditions}{7}
\newcommand{\NumSeeds}{3}
\newcommand{\NumTrainPerCondition}{3{,}009}
\newcommand{\NumHeldoutPerStudent}{561}
\newcommand{\NumEvalSamples}{35{,}343}
\newcommand{\TeacherStrong}{90.93}
\newcommand{\TeacherStrongUp}{60.37}
\newcommand{\TeacherSoft}{17.85}
\newcommand{\TeacherSoftUp}{11.93}
\newcommand{\TeacherStyleCtrl}{17.05}
\newcommand{\TeacherBaselineStrong}{0.20}
\newcommand{\TeacherBaselineStyleCtrl}{1.20}
\newcommand{\TeacherSoftCompliance}{91.45}
\newcommand{\TeacherStyleCtrlCompliance}{83.15}
\newcommand{\TeacherStrongRetention}{66.4}
\newcommand{\TeacherSoftRetention}{66.8}
\newcommand{\TeacherSelfRetention}{\TeacherStrongRetention}
\newcommand{\GemmaStrongTransfer}{88.9}
\newcommand{\OlmoStrongTransfer}{80.9}
\newcommand{\QwenStrongTransfer}{45.2}
\newcommand{\GemmaRrelMean}{89}
\newcommand{\GemmaRrelSD}{23}
\newcommand{\OlmoRrelMean}{112}
\newcommand{\OlmoRrelSD}{16}
\newcommand{\QwenRrelMean}{65}
\newcommand{\QwenRrelSD}{30}
\newcommand{\OlmoSoftRate}{31.99}
\newcommand{\OlmoSoftStd}{3.04}
\newcommand{\OlmoSoftTransfer}{179.2}

\newcommand{\OlmoSoftSeparation}{49}
\newcommand{\QwenSoftSeparation}{9.6}
\newcommand{\PilotN}{452}
\newcommand{\PilotBaselineK}{23}
\newcommand{\PilotStrongK}{450}
\newcommand{\PilotBaselineRate}{5.09}
\newcommand{\PilotStrongRate}{99.56}
\newcommand{\PilotBaselineCI}{$[3.41, 7.52]$}
\newcommand{\PilotStrongCI}{$[98.40, 99.88]$}
\newcommand{\PilotGapPP}{94.47}
\newcommand{\PilotGapCI}{$[92.35, 96.59]$}

\newcommand{\KappaNumAnnotators}{3}
\newcommand{\KappaNumItems}{200}
\newcommand{\StrongKappa}{0.84}
\newcommand{\StrongKappaCI}{[0.72,\,0.94]}
\newcommand{\StyleKappa}{0.78}
\newcommand{\StyleKappaCI}{[0.64,\,0.90]}
\newcommand{\PooledKappa}{0.81}
\newcommand{\StrongFleiss}{0.88}
\newcommand{\StyleFleiss}{0.81}

\newcommand{\StrongJudgePrec}{94}
\newcommand{\StrongJudgeNPV}{90}
\newcommand{\StyleJudgePrec}{86}
\newcommand{\StyleJudgeNPV}{92}

\newcommand{\StrongReweighted}{0.76}
\newcommand{\StyleReweighted}{0.40}
\newcommand{\HFivePilotN}{5}
\newcommand{\HFivePilotBase}{3.00}
\newcommand{\HFivePilotSTwenty}{3.20}
\newcommand{\HFivePilotSFifty}{4.20}
\newcommand{\HFivePilotSCTwenty}{3.00}
\newcommand{\HFivePilotSCFifty}{3.20}

\InputIfFileExists{sections/generated_numbers.tex}{}{}

\vspace{-2mm}
\title{Asking Back: Interaction-Layer\\ Antidistillation Watermarks}

\vspace{-2mm}
\author{%
  Guang Yang$^{1}$, Amir Ghasemian$^{1}$,Fengchen Liu$^{2}$, Zhong Wang$^{2}$,\\
   Ninareh Mehrabi$^{3}$, Homa Hosseinmardi$^{1}$ \\
  $^{1}$University of California, Los Angeles,
  $^{2}$Lawrence Berkeley National Laboratory,
  $^{3}$Meta%
}

\begin{document}
\maketitle

\input{sections/01_abstract.tex}
\newpage
\input{sections/02_introduction.tex}
\input{sections/03_related.tex}
\input{sections/03b_pilot.tex}
\input{sections/04_method.tex}

\input{sections/05_setup.tex}
\input{sections/06_results.tex}

\input{sections/07_discussion.tex}

\input{sections/09_conclusion.tex}

\bibliographystyle{plainnat}
\bibliography{references}

\input{sections/11_appendix.tex}

% \newpage
% NeurIPS Paper Checklist (required for submission).
% \answerYes / \answerNo / \answerNA / \answerTODO are pre-defined by neurips_2026.sty.
% \input{checklist.tex}

\end{document}

%% file: sections/generated_numbers.tex
% Auto-checkable numerics surfaced as macros so the prose, abstract,
% headline table, and figures all read from one place.
%
% Source files (verified 2026-04-29 against on-disk artifacts):
%   source/fingerprints/teacher/teacher_fingerprint_rates.json
%   source/fingerprints/v1_run/fingerprint_rates.json
%   source/data/teacher/{baseline,strong,soft,style_control,
%                        baseline_up,soft_up,strong_up}.jsonl
%   source/results/{qwen,gemma,olmo}/*_seed{42,1815,7026}_heldout1000.jsonl
%
% If you change a number here, also update tables/table2_headlines.tex,
% paper/MISSING_DATA.md, and the abstract.

% --- experimental scope ---
\renewcommand{\NumStudents}{63}                     % 3 families * 7 conds * 3 seeds
\renewcommand{\NumFamilies}{3}
\renewcommand{\NumConditions}{7}
\renewcommand{\NumSeeds}{3}
\renewcommand{\NumTrainPerCondition}{3{,}009}
\renewcommand{\NumHeldoutPerStudent}{561}
\renewcommand{\NumEvalSamples}{35{,}343}            % 3 * 7 * 3 * 561

% --- teacher reference (Llama-3.3-70B-Instruct; n=3009 / cond) ---
\renewcommand{\TeacherStrong}{90.93}
\renewcommand{\TeacherStrongUp}{60.37}
\renewcommand{\TeacherSoft}{17.85}
\renewcommand{\TeacherSoftUp}{11.93}
\renewcommand{\TeacherStyleCtrl}{17.05}
\renewcommand{\TeacherBaselineStrong}{0.20}
\renewcommand{\TeacherBaselineStyleCtrl}{1.20}

% --- teacher compliance rates on the soft / style-control subsets
% (judge-measured rate of the marker on the 582-row 20% mix)
\renewcommand{\TeacherSoftCompliance}{91.45}
\renewcommand{\TeacherStyleCtrlCompliance}{83.15}

% --- two-level paraphrase robustness (teacher-self) ---
\renewcommand{\TeacherStrongRetention}{66.4}        % 60.37 / 90.93
\renewcommand{\TeacherSoftRetention}{66.8}          % 11.93 / 17.85

% --- relative transfer rates (student / teacher), strong condition ---
\renewcommand{\GemmaStrongTransfer}{88.9}           % 80.87 / 90.93
\renewcommand{\OlmoStrongTransfer}{80.9}            % 73.56 / 90.93
\renewcommand{\QwenStrongTransfer}{45.2}            % 41.08 / 90.93

% --- student-relative paraphrase retention R_rel = R_S/R_T on STRONG.
% Means propagated by the delta method from per-seed strong/strong_up
% rates in Table 7 against the teacher reference R_T = 60.37/90.93.
\renewcommand{\GemmaRrelMean}{89}
\renewcommand{\GemmaRrelSD}{23}
\renewcommand{\OlmoRrelMean}{112}
\renewcommand{\OlmoRrelSD}{16}
\renewcommand{\QwenRrelMean}{65}
\renewcommand{\QwenRrelSD}{30}

% --- amplification effect (OLMo soft) ---
\renewcommand{\OlmoSoftRate}{31.99}
\renewcommand{\OlmoSoftStd}{3.04}
\renewcommand{\OlmoSoftTransfer}{179.2}             % 31.99 / 17.85

% --- soft-rate fold-over-baseline multipliers (H4)
% These are rate-over-baseline-mean ratios, not std-dev separations.
% Rendered with a "x" suffix in the prose; Gemma's baseline is 0
% across three seeds, so the multiplier is unbounded.
         % 15.52 / 0.00 (baseline = 0)
\renewcommand{\OlmoSoftSeparation}{49}              % 31.99 / 0.65
\renewcommand{\QwenSoftSeparation}{9.6}             % 15.98 / 1.67

% --- pilot study (Phase 1, within-family Qwen3.5-9B -> 0.8B)
% Source: paper/result_v1/artifacts/generations/cleaned/qmark_rate.json
% Computed deterministically by paper/result_v1/{clean_pilot,count_qmark_rate}.py
% on the pairwise-cleaned heldout-1000 outputs.
\renewcommand{\PilotN}{452}
\renewcommand{\PilotBaselineK}{23}
\renewcommand{\PilotStrongK}{450}
\renewcommand{\PilotBaselineRate}{5.09}
\renewcommand{\PilotStrongRate}{99.56}
\renewcommand{\PilotBaselineCI}{$[3.41, 7.52]$}
\renewcommand{\PilotStrongCI}{$[98.40, 99.88]$}
\renewcommand{\PilotGapPP}{94.47}
\renewcommand{\PilotGapCI}{$[92.35, 96.59]$}

% --- legacy macros from v1 draft (kept for backwards compatibility) ---

% --- judge--human kappa validation (Appendix G; n=200 verdict-stratified
% items, 3 fresh annotators, judge vs. 3-rater majority).
% Source: human_test/task2_judge_kappa/task2/task_final/final_ana/
\renewcommand{\KappaNumAnnotators}{3}
\renewcommand{\KappaNumItems}{200}
\renewcommand{\StrongKappa}{0.84}
\renewcommand{\StrongKappaCI}{[0.72,\,0.94]}
\renewcommand{\StyleKappa}{0.78}
\renewcommand{\StyleKappaCI}{[0.64,\,0.90]}
\renewcommand{\PooledKappa}{0.81}
\renewcommand{\StrongFleiss}{0.88}
\renewcommand{\StyleFleiss}{0.81}

\renewcommand{\StrongJudgePrec}{94}
\renewcommand{\StrongJudgeNPV}{90}
\renewcommand{\StyleJudgePrec}{86}
\renewcommand{\StyleJudgeNPV}{92}

\renewcommand{\StrongReweighted}{0.76}    % at 22.3\% natural yes-rate
\renewcommand{\StyleReweighted}{0.40}     % at 3.7\% natural yes-rate

% --- H5 in-lab study (N=20, 5x5 Latin-square balanced).
% Source: human_test/h5_0506/data_extracted/data/ (20 of 29 raw
% participants completed all 5 sessions).
% Detection macros (BaseDet/STwentyDet/...) intentionally not
% redefined here: the loose any-non-empty-freetext rate captures
% session-comment habit (33% at baseline) more than marker-specific
% detection, so the paper no longer reports these.
\renewcommand{\HFivePilotN}{20}
\renewcommand{\HFivePilotBase}{3.55}
\renewcommand{\HFivePilotSTwenty}{3.45}
\renewcommand{\HFivePilotSFifty}{3.55}
\renewcommand{\HFivePilotSCTwenty}{3.70}
\renewcommand{\HFivePilotSCFifty}{3.75}

%% file: sections/01_abstract.tex
% Abstract --- ~200 words. Headline numbers only; per-sub-result numbers live in §6.
\begin{abstract}
    This work is motivated in part by a parallel genome-model study (\textsc{GenoTrace} on \texttt{GenomeOcean}~\citep{zhou2025genomeocean}; App.~\ref{app:genotrace}), where a token-level distillation watermark suggested a broader question: can a trace survive distillation when its carrier is not token statistics, but interaction behavior?
    Detecting unauthorized knowledge distillation from a deployed language-model API is hard precisely because the defender controls
    neither the attacker's training pipeline nor, in the API setting,
    the next-token logits. Existing defenses operate on the teacher's output tokens, either by biasing the next-token distribution (green-list watermarks, cryptographic schemes, antidistillation sampling) or by rewriting outputs after generation (defensive output rewriting). Recent work shows that an attacker who paraphrases harvested responses before training can strip these signals without losing the underlying knowledge---defeating the watermark while preserving the distillation.
    %Existing defenses embed the trace inside the token stream---green-list watermarks, cryptographic schemes, antidistillation sampling, defensive output rewriting---and recent work shows these signals can be detached by paraphrasing or pre-distillation rewriting while the underlying knowledge transfer is preserved. 
    We propose \emph{interaction-layer antidistillation
    watermarks}, which move the trace one layer higher, into the
    teacher's interaction \emph{behavior}: the defender wraps the
    teacher with a system prompt that intermittently induces a
    behavioral marker---an explicit follow-up question, a low-frequency variant of it, or a declarative restatement. An oblivious distiller inherits the behavior in their student, and the
    defender later audits through black-box queries with a human-validated LLM-as-judge (Cohen's $\kappa{=}\StrongKappa$/$\StyleKappa$ on strong/style rubrics). Across $\NumStudents$
    LoRA-distilled students under a \texttt{Llama-3.3-70B-Instruct}
    teacher ($\NumEvalSamples$ judged samples), behavioral watermarks
    transfer at $\GemmaStrongTransfer\%$ (Gemma) / $\OlmoStrongTransfer\%$ (OLMo) / $\QwenStrongTransfer\%$ (Qwen)
    relative fidelity (H1, H2); under non-adaptive DIPPER
    prompt paraphrasing, robustness decomposes into a teacher-self ceiling
    ($\sim\!\TeacherSelfRetention\%$) and a student-relative retention of
    $21$--$112\%$, with one family (OLMo) preserving the watermark
    \emph{above} the teacher itself (H3, F-Amp);
    low-density ($\sim\!20\%$) explicit and implicit declarative
    variants both transfer reliably above per-family baseline (H4,
    F-Style). An $N{=}\HFivePilotN$ in-lab study under a
    pre-registered Latin-square protocol shows all marker variants
    within $0.22$ Likert step of baseline; TOST, Friedman, and
    Bonferroni-Wilcoxon all support H5.
    %Judge--human Cohen's $\kappa{=}\StrongKappa$/$\StyleKappa$ on strong/style rubrics. 
    The interaction layer is, on present evidence,
    a viable design locus for antidistillation watermarking,
    complementary to token-, model-, and reasoning-trace-layer defenses.
    \end{abstract}

%% file: sections/02_introduction.tex
\section{Introduction}
\label{sec:intro}

Frontier language models are increasingly served through paid black-box
APIs, and unauthorized knowledge distillation against those APIs is now a
mature commercial threat that has motivated a range of antidistillation\footnote{We use \emph{antidistillation} in the broader
sense of \citet{xu2026adfingerprint}---making unauthorized
distillation auditable through a downstream-detectable
mark---rather than the strict sense of
\citet{savani2025antidistillation}, where the teacher's outputs are crafted to deliberately degrade the
resulting student.} defenses~\citep{savani2025antidistillation,pan2025watermarks,
li2025doge}. An adversary harvests prompt--response pairs from the teacher,
fine-tunes a smaller student, and ships the student without permission.
The defender's central question is therefore not whether harvesting can be
\emph{blocked}---it cannot, in any deployable sense---but whether the
defender can leave an \emph{auditable trace} that survives distillation
and that can later be verified through black-box queries to the suspected
student.

Existing defenses are organized by the layer at which the trace is
embedded. The \textbf{token-distribution layer} family
\citep{kirchenbauer2023watermark,aaronson2023watermark,
christ2024undetectable,savani2025antidistillation} perturbs the
teacher's next-token distribution. The \textbf{model layer} family
\citep{xu2024instructional,li2025doge} edits the teacher itself so
that the harvested text carries a model-side mark. The
\textbf{reasoning-trace layer} family
\citep{ma2026tracerewriting,ding2025part} reorders or rewrites
chain-of-thought traces. All three share an attack surface operating
on the \emph{tokens} the attacker harvests, and recent
removal-attack work~\citep{pan2025watermarks,krishna2023paraphrasing,
sadasivan2024canai,an2026ditto} shows that pre-distillation rewriting
can detach a student from the inherited signal while preserving the
knowledge transfer. The deeper structural gap is that even when a
surface-text signal survives, it does not specify a distinctive,
queryable \emph{behavior} in the distilled student.

Our starting point was a parallel genome-model study, where a
KGW-style logit-bias watermark~\citep{kirchenbauer2023watermark}
was ported into the \texttt{GenomeOcean}~\citep{zhou2025genomeocean}
LoRA distillation pipeline and retained the radioactive-data
property of~\citep{sander2024radioactive} in a non-text regime
(\textsc{GenoTrace}; App.~\ref{app:genotrace}). For conversational
LLMs, the analogous carrier need not remain at the token layer:
the defender also controls the interaction frame between user and
model. That observation led to the asking-back markers studied here.
We return to this connection in \S\ref{sec:f-modality}.

\begin{figure}[t]
    \centering
    \includegraphics[width=0.89\linewidth]{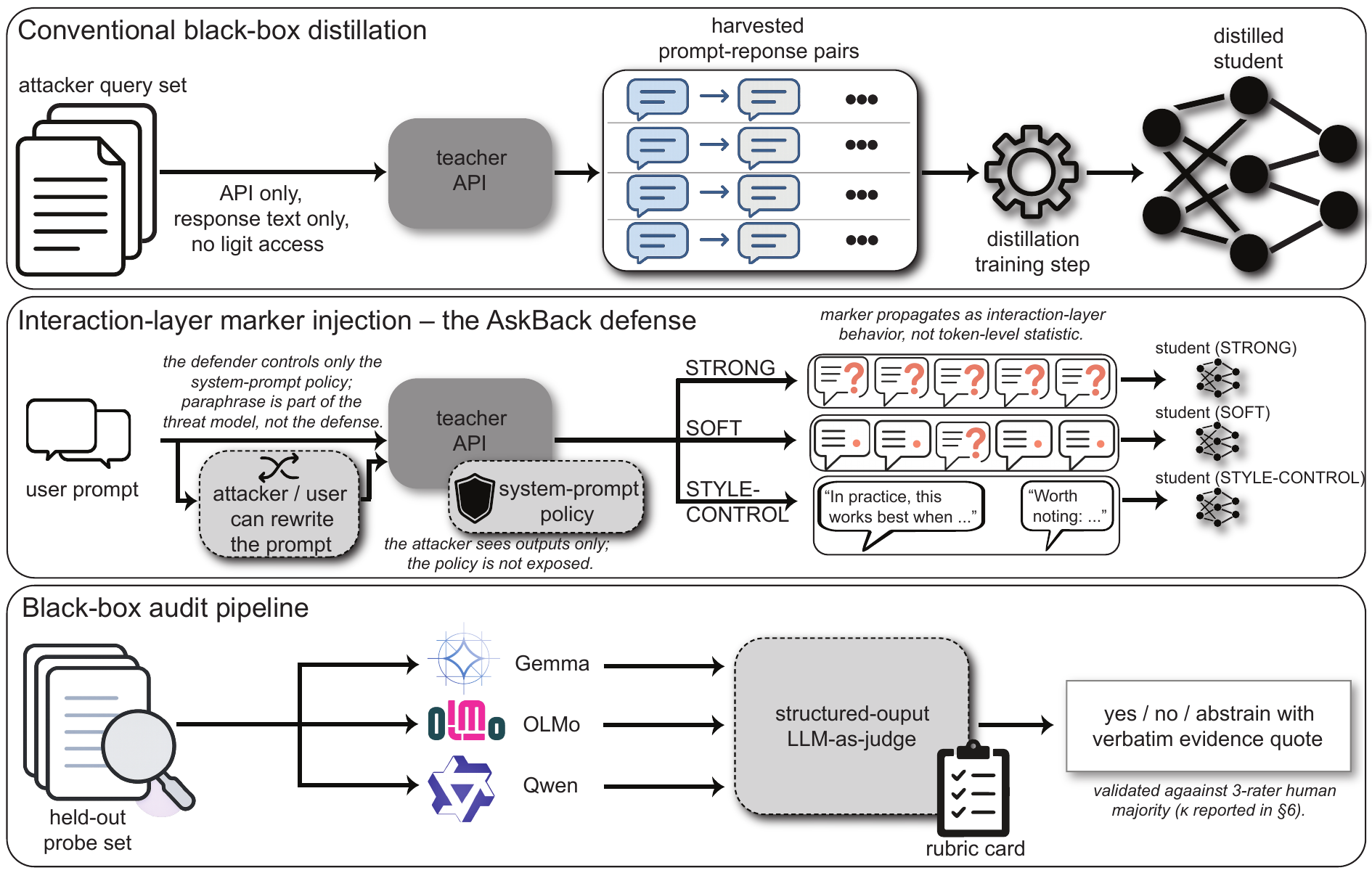}
    \caption{The three layers of the \AskBack{} threat model.
    \emph{Top:} the attacker harvests prompt--response pairs from the
    teacher API; the defender has no control over which prompts.
    \emph{Middle:} the defender wraps the teacher with a system prompt that injects a behavioral marker (asking-back, soft, or style-control) at a chosen density~$\rho$; harvested pairs carry the marker into the student. 
%the defender wraps the teacher with one of seven system-prompt conditions (section~\ref{sec:setup}), three of which intermittently inject the behavioral marker at a chosen density~$\rho$. The mixture is aligned by prompt ID across conditions, so the soft and style-control datasets are identical to the baseline dataset on $\sim\!80\%$ of rows. 
    \emph{Bottom:} the defender audits the suspect student
    through black-box queries with an LLM-as-judge whose rubric is
    blind to family and condition, and computes the relative transfer
    rate $\tau_{\mathrm{rel}}(S, B) = \tau(S, B) / \tau(T, B)$.}
    \label{fig:pipeline}
\end{figure}

In this work, we propose a fourth, complementary locus---the \emph{interaction layer} (Figure~\ref{fig:pipeline}).
The defender wraps the teacher with a system prompt that intermittently
induces a behavioral marker $B$: \emph{asking-back} (a follow-up
question), or its implicit cousins (a low-density variant, or a
declarative restatement of the user's request). An oblivious distiller who collects teacher outputs without access to the hidden trigger policy
trains a student on a corpus in which $B$ appears at some defender-chosen
density~$\rho$. Three properties make this useful: $B$ requires
\emph{no logit access} (the trace lives in the response text);
it is \emph{distillation-transferable} because $B$ is part of the next
tokens the student is asked to imitate; and it operates at a
\emph{semantic granularity} partially insensitive to prompt-side
paraphrasing (response-side rewriting is left to future work, Appendix, section~\ref{app:lim}).

%\paragraph{Hypotheses.}
We organize the experiments around five hypotheses any deployable
interaction-layer watermark\footnote{We use \emph{watermark} throughout in an operational
sense---a detectable behavioral bias deliberately induced by the
defender---rather than in the cryptographic or information-theoretic
sense of an undetectable or provably secure scheme.} must satisfy:
\textbf{H1--Learnability}: a student trained on watermarked teacher
outputs exhibits $B$ at a rate substantially above its
baseline-trained counterpart;
\textbf{H2--Cross-family generalization}: the effect holds across
pretrained model families with different architectures,
vocabularies, and post-training intensities;
\textbf{H3--Paraphrase robustness}: the watermark partially
survives a non-adaptive prompt-side paraphrase (DIPPER on the
user prompt), decomposed into a \emph{teacher-self} ceiling and
a \emph{student-relative} component;
\textbf{H4--Density--detectability trade-off}: even when $B$
appears in only $\sim\!20\%$ of training rows, the student-side
detection rate is reliably above the per-family noise floor;
\textbf{H5--Stealth}: a $\sim\!20\%$-density watermark is not
perceived by users as a clear degradation of interaction quality.

%\paragraph{Two-phase empirical structure.}
Our evaluation has two phases:
%\textbf{Phase 1} (\S\ref{sec:pilot}), 
a within-family
\texttt{Qwen3.5-9B-Instruct}~$\to$~\texttt{Qwen3.5-0.8B-Base}
case study at $N{=}\PilotN$ paired prompts under a deterministic
regex (judge-free) %; strong-vs.-baseline gap $+\PilotGapPP\,$pp);
and %\textbf{Phase 2} (\S\ref{sec:method}--\S\ref{sec:results}), 
a $\NumFamilies\!\times\!\NumConditions\!\times\!\NumSeeds=\NumStudents$-student
matrix under a
\texttt{Llama-3.3-70B-Instruct}~\citep{grattafiori2024llama3}
teacher and a structured-output
\texttt{gpt-oss-120b}~\citep{gptoss2025} judge
($\NumEvalSamples$ judged samples). The three markers and the three
student families are also described in sections~\S\ref{sec:formulation} and \S\ref{sec:setup}.

%\paragraph{Contributions.}
Our contributions are threefold. \textbf{(i)~Problem framing.} We formalize %position 
the \emph{interaction
layer} %as a fourth, complementary locus and formalize the
watermark (Definition~\ref{def:watermark}) and three metrics: $\tau,\tau_{\mathrm{rel}},R$
(Definitions~\ref{def:transfer}--\ref{def:robust},
Appendix, section~\ref{app:defs}). These are the behavioral counterparts of
green-list~\citep{kirchenbauer2023watermark} and
radioactive-data~\citep{sablayrolles2020radioactive} metrics. 
\textbf{(ii)~Scale.} To our knowledge the broadest cross-family
distillation watermarking study:
$\NumFamilies\!\times\!\NumConditions\!\times\!\NumSeeds=\NumStudents$
LoRA students, $\NumEvalSamples$ judged samples, single shared
rubric.
\textbf{(iii)~Two-level paraphrase robustness.} Paraphrase
decomposes into a teacher-self ceiling
($\sim\!\TeacherStrongRetention\%$ under
DIPPER~\citep{krishna2023paraphrasing}) and a student-relative
factor, with one family preserving the watermark above the
teacher itself.%; this reframes the standard one-number robustness narrative.

%% file: sections/03_related.tex
\section{Related Work}
\label{sec:related}

A growing body of work studies how to make unauthorized distillation
auditable by embedding detectable signals into the teacher's outputs.
These approaches differ primarily in where the signal is introduced
and how it survives downstream processing. Here, we organize prior antidistillation work by the layer at which the
defender's signal is embedded. %; our contribution is to add a fourth.

%\paragraph{Token-distribution layer.}
Early work focuses on the token distribution. 
Green-list watermarks~\citep{kirchenbauer2023watermark} bias generation toward a hashed subset of tokens, while cryptographic
schemes~\citep{aaronson2023watermark} use keyed pseudorandom mappings,
and undetectable schemes~\citep{christ2024undetectable} aim for stronger
formal guarantees.
%Green-list watermarks~\citep{kirchenbauer2023watermark} bias logitstoward a hashed subset; cryptographic schemes~\citep{aaronson2023watermark} use keyed pseudorandom mappings; undetectable schemes~\citep{christ2024undetectable} attain a stronger formal guarantee; 
Antidistillation sampling~\citep{savani2025antidistillation} instead modifies the distribution to degrade its usefulness for downstream learners.
However, these methods typically depend on \emph{logit access} or
surface-text statistics, making them vulnerable to paraphrasing and
rewriting attacks~\citep{krishna2023paraphrasing,sadasivan2024canai}.
Empirical work~\citep{pan2025watermarks,an2026ditto} shows that such
signals can weaken or be removed entirely under realistic attacker
pipelines.
%antidistillation sampling~\citep{savani2025antidistillation} steers the same distribution to be useless to a downstream learner. \citet{pan2025watermarks} report non-trivial signal loss when these survive distillation, and spoofing-style attacks~\citep{an2026ditto} demonstrate the inverse vulnerability. 
%All rely on \emph{logit access} or surface-text statistics that paraphrasing~\citep{krishna2023paraphrasing,sadasivan2024canai} attacks directly.

%\paragraph{Model layer.}
A complementary line of work embeds signals at the model level.
Instructional fingerprinting~\citep{xu2024instructional} introduces
model-side signatures through fine-tuning, while defensive output
generation~\citep{li2025doge} modifies the final layer of the teacher.
%Instructional fingerprinting~\citep{xu2024instructional} bakes a model-side fingerprint into weights via fine-tuning; defensive output generation~\citep{li2025doge} adapts the teacher's final layer.
Related approaches extend these ideas to distillation-aware
fingerprinting~\citep{xu2026adfingerprint}. 
%\citet{xu2026adfingerprint} extend this to distillation-aware token-layer fingerprinting. 
These methods generally assume that the defender has control over the teacher model and can modify its parameters or output layer.
%The shared assumption is that the defender owns the teacher and is willing to modify it.
%\paragraph{Reasoning-trace layer.}
Other approaches operate on reasoning traces. Some have tried
trace rewriting~\citep{ma2026tracerewriting} or, for example, PART reorders sub-conclusions to preserve answer
information while disrupting inferential structure~\citep{ding2025part}. 
%\citet{ma2026tracerewriting} rewrite reasoning traces;PART~\citep{ding2025part} reorders sub-conclusions to preserve answer information while disrupting inferential structure. 
These approaches assume the teacher exposes (and controls) a reasoning trace, which restricts them to specific deployment patterns.

%\paragraph{Distillation provenance and behavioral attribution.}
More broadly, provenance and attribution have been studied through
techniques such as membership inference~\citep{shokri2017membership,mireshghallah2022memorization},
though these often require impractically large query budgets in API
settings.
% Membership inference~\citep{shokri2017membership,mireshghallah2022memorization}requires query budgets impractical at API scale. 
Radioactive data~\citep{sablayrolles2020radioactive} is the closest classical analogue: a defender perturbs training data so that a downstream learner's parameters carry a detectable mark. 
Our approach adopts a similar philosophy of embedding a signal in the data and detecting it through the trained model, but operates at the level of behavior rather than parameters, avoiding
the need for white-box access. 
%We carry the same ``mark in the data, detect via the downstream artifact'' philosophy at the behavioral rather than the parameter layer, sidestepping the need for white-box access. 
Prior work also observes that models can exhibit incidental stylistic
fingerprints arising from training~\citep{sun2024fingerprint}; in contrast, 
%\citet{sun2024fingerprint} observe \emph{incidental} stylistic fingerprints arising from training; 
our watermarks are \emph{engineered} signals deliberately induced by the
defender via system prompts, so the two are complementary (incidental fingerprints might be used to corroborate, but cannot substitute for, an engineered watermark). 
The underlying clarifying-question behavior is itself a studied dialogue
capability~\citep{andukuri2025clarifying}, where our novelty here is the
\emph{defensive} use of such a behavior as a transferable, auditable marker. 
Concurrent work~\citep{cloud2026subliminal} documents that distillation can transmit teacher behavioral traits even through training data with no language-level content, a parameter-space
channel that operates only between teacher and student sharing
initialization; we discuss the boundary between this regime and
ours in \S\ref{sec:discussion}.

%\paragraph{Position of this work.}
%The four loci are \emph{not in competition}. Token-distribution signals fail under response-level paraphrasing; behavioral signals fail under prompt-level rewriting. Because the two attack surfaces are disjoint, signals placed at different loci are best understood as complementary primitives rather than as alternatives---each adds an audit channel that is not redundant with the others.
Taken together, these lines of work suggest that different classes of
signals occupy distinct points in the design space. Token-level
approaches are vulnerable to response-level rewriting, while
behavioral signals can be disrupted by prompt-level transformations.
Rather than viewing these approaches as competing, we treat them as
complementary: each provides an additional audit channel with a
different failure mode. Our work builds on this perspective by
focusing on interaction-level behavior as a distinct and practical
locus for antidistillation watermarking.

%% file: sections/03b_pilot.tex
%\section{Pilot Study (Phase 1)}
\section{Overview of Approach}
\subsection{Within-family case study}  
\label{sec:pilot}

%Phase 1 uses a within-family \texttt{Qwen3.5-9B-Instruct}~$\to$~\texttt{Qwen3.5-0.8B-Base} pair, the strong condition only, a single seed, and a deterministic question-mark-ending regex on the cleaned response field (no LLM-as-judge).  Setup, the pairwise sentence-ending cleaning filter, and the regex specification are in Appendix~\ref{app:pilot}. 
We begin with a controlled, within-family experiment to establish whether interaction-layer markers can survive distillation under the most favorable conditions. Concretely, we distill from a \texttt{Qwen3.5-9B-Instruct} teacher into a \texttt{Qwen3.5-0.8B-Base} student using only the strong condition. The experiment uses a single seed and evaluates on 452 paired prompts. We adopt a strict proxy for the behavioral marker: whether the response ends in a question mark, detected via a deterministic regex applied to cleaned outputs (see Appendix, section~\ref{app:pilot} for regex specification).
% Because the regex requires a literal \texttt{?}-ending it is a \emph{strict lower bound} on the broader behavioral marker measured by the \texttt{gpt-oss-120b} judge in Phase 2; absolute numbers from the two phases are therefore not directly comparable.
This measurement is intentionally conservative, requiring a literal \texttt{?} at the end of the response, to capture only the most explicit form of the asking-back behavior, and it should be interpreted as a lower bound relative to the broader semantic marker used later with the \texttt{gpt-oss-120b} judge. %Absolute values are therefore not directly comparable to Phase 2.

%\paragraph{Result: H1 supported within-family at strict measurement.}
%\label{sec:pilot-result}
We find that the student trained on watermarked data produces question-ending responses at a rate of $\PilotStrongRate\%$ ($\PilotStrongK/\PilotN$, Wilson 95\% CI \PilotStrongCI), whereas the baseline-trained student does so only $\PilotBaselineRate\%$ of the time ($\PilotBaselineK/\PilotN$, CI \PilotBaselineCI), yielding a gap of $+\PilotGapPP$ percentage points (CI \PilotGapCI) that excludes zero by a wide margin, showing the behavior is learned almost deterministically. 
%On the $\PilotN$ paired prompts, the strong rate is $\PilotStrongRate\%$ ($\PilotStrongK/\PilotN$, Wilson 95\% CI \PilotStrongCI) vs.\ baseline $\PilotBaselineRate\%$ ($\PilotBaselineK/\PilotN$, CI \PilotBaselineCI), a $+\PilotGapPP\,$pp gap (CI \PilotGapCI) excluding zero by a wide margin.  %: the student follows the ``end with \texttt{?}'' instruction on $450/452$ prompts, while the baseline does so on only $23/452$. 
This provides strong evidence for H1 (\emph{learnability through knowledge distillation}, KD~\citep{hinton2015distilling}) in the within-family setting, even under a deliberately strict proxy  
%Per-condition $k/N$ and Wilson CIs are reported in 
(see Table~\ref{tab:pilot}, Appendix, section~\ref{app:pilot}, for more details).
%This supports H1 (\emph{learnability through KD}) in the within-family setting at the strictest available measurement. The strong rate is near the ceiling: the student obeys ``end with \texttt{?}'' on $450/452$ paired prompts; the baseline does so on only $23/452$. Per-condition $k/N$ and Wilson CIs are tabulated in Table~\ref{tab:pilot} (Appendix~\ref{app:pilot}).
%\paragraph{Lessons motivating Phase 2.}
%\label{sec:pilot-lessons}
%The pilot leaves H2--H5 untested and surfaces four design constraints that Phase 2 adopts: \textbf{(i)~base/PT student weights} (so an inherited assistant prior cannot be mistaken for the distilled marker); \textbf{(ii)~ID-aligned mixing of low-density variants} (so any rate difference between soft, soft\_up, style\_control is attributable to the marker, not to prompts); \textbf{(iii)~a separate $70$B-class teacher} decoupling cross-family results from a same-family vocabulary or post-training pipeline; and \textbf{(iv)~a structured-output \llmjudge{}}, because regex cannot detect the soft and style-control markers (not syntactically interrogative). Phase 2 (\S\ref{sec:method}--\S\ref{sec:results}) instantiates these on a $\NumStudents$-student matrix.
This design is limited in scope and does not address cross-family generalization, robustness to paraphrasing, or behavior at lower marker densities, and it relies on a proxy that captures only one surface form of the behavior. In the following, we adopt (i) base/pretrained (PT) student checkpoints to avoid conflating inherited assistant priors with the distilled marker; (ii) ID-aligned mixing for low-density variants, so that differences across conditions are attributable to the marker rather than the prompts; (iii) a separate 70B-class teacher, decoupling the results from within-family artifacts; and (iv) a structured-output \llmjudge{}, which allows us to detect both explicit and implicit forms of the behavior. We turn to this broader evaluation next (sections \S\ref{sec:formulation}--\S\ref{sec:results}).

%% file: sections/04_method.tex
\subsection{Formalization}
\label{sec:formulation}

%\subsection{Threat model and watermark transfer metrics}
 
We formalize the watermark at the level of interaction. Rather than modifying
token probabilities or model parameters, the defender prepends a system prompt
$s_B$ to a teacher $T$, producing a wrapped teacher $T_{s_B}$ whose responses
can be tested for a \emph{behavioral marker} $B$. A student trained on
$T_{s_B}$'s outputs inherits the marked behavior through standard distillation.
%The watermark lives at the \emph{interaction layer}: the defender wraps a teacher $T$ with a system prompt $s_B$ that induces a behavioral marker $B:\mathcal{P}\!\times\!\mathcal{Y}\!\to\!\{0,1\}$ at density $\rho\in[0,1]$ (Definition~\ref{def:watermark}); the student inherits that behavior through standard distillation on the teacher's outputs.

\iffalse
\begin{definition}[Interaction-layer behavioral watermark]
\label{def:watermark}
For teacher $T$ and prompt distribution $\mathcal{P}$, an
\emph{interaction-layer behavioral watermark} is a tuple
$(B,s_B,\rho)$ where $s_B$ is a system prompt that, prepended to
$T$, induces marker rate
$\rho\!=\!\mathbb{E}_{p\sim\mathcal{P}}[B(p,T_{s_B}(p))]\!\in\![0,1]$.
\end{definition}
\fi
\begin{definition}[Interaction-layer behavioral watermark]
\label{def:watermark}
For teacher $T$, prompt space $\mathcal{X}$, response space $\mathcal{Y}$,
and prompt distribution $\mathcal{P}$ over $\mathcal{X}$, an
\emph{interaction-layer behavioral watermark} is a triple $(B, \Pi_B, \rho)$
where $B\colon \mathcal{X}\times\mathcal{Y}\to\{0,1\}$ is a marker function,
$\Pi_B$ is a (possibly stochastic) \emph{trigger policy} that selects a
system prompt $s_B \sim \Pi_B(p)$ for each input $p$, and $\rho\in[0,1]$ is
the induced marker rate:
\[
\rho \;=\; \mathbb{E}_{p \sim \mathcal{P},\; s_B \sim \Pi_B(p)} \big[ B(p,\, T_{s_B}(p)) \big].
\]
\end{definition}

$T_{s_B}(\cdot\mid p)$ denotes the response distribution of $T$ when
$s_B$ is prepended to $p$ as a system prompt. The constant-prompt case
$\Pi_B(p){=}\delta_{s_B}$ recovers the standard \textsc{Strong}
configuration; ID-deterministic mixtures
(\textsc{Soft}, \textsc{Style-control}) instantiate
$\Pi_B$ as a two-prompt policy gated by an indicator
$\mathbf{1}\{p\in\mathcal{I}_{20}\}$.

We reserve \emph{watermark} for the engineered signal $(B, \Pi_B, \rho)$ and
\emph{fingerprint} for incidental stylistic regularities that arise without
intervention; only watermarks are calibrated and auditable. 
%We reserve \emph{watermark} for the engineered signal $(B,s_B,\rho)$ and \emph{fingerprint} for incidental stylistic regularities a model exhibits without defender intervention~\citep{sun2024fingerprint}; only watermarks are calibrated and auditable.
%\paragraph{Transfer and robustness.}
A student $S$ trained on $T_{s_B}$'s outputs and queried without any system prompt on a held-out distribution $\mathcal{P}_\text{held}$ defines absolute and relative
transfer rates $\tau(S,B)$ and $\tau_\text{rel}(S,B)$ (see Definition~\ref{def:transfer}, Appendix, section~\ref{app:defs}).
%A student $S$ trained on harvested teacher outputs and evaluated without any system prompt on a disjoint $\mathcal{P}_{\mathrm{held}}$ defines the absolute and relative transfer rates $\tau(S,B),\,\tau_{\mathrm{rel}}(S,B)$ (Def.~\ref{def:transfer}). 
Under a prompt-side paraphraser $\pi$
(applied to the user prompt, not to the teacher's response),
robustness decomposes into a \emph{teacher-self} component $R_T$
(the teacher's marker rate under prompt-side input shift, normalized by its clean rate) and a \emph{student-relative} component $R_S$
%(student marker rate on the \emph{clean} held-out set when trained on paraphrased data
(the student's marker rate on the clean held-out set when trained on paraphrased data, normalized by its rate when trained on clean data; the two components live on different distributions by design, see Definition~\ref{def:robust}, Appendix, section~\ref{app:defs}):
\begin{equation}
\label{eq:two-level}
R_S \;=\; R_T\cdot R_{\mathrm{rel}},\qquad
R_{\mathrm{rel}}(B,\pi)=R_S/R_T.
\end{equation}
Full formal definitions, finite-sample estimators, and the
calibration-invariance argument that lets us compare students and
teacher under one shared judge are provided in Appendix, section~\ref{app:defs}.
$R_{\mathrm{rel}}>1$ would mean the student preserves the
watermark more reliably than the teacher itself, a regime we
observe on one family (F-Amp, section \S\ref{sec:discussion}).

%\paragraph{Detection.}
In practice, $B$ is implemented by a structured-output \llmjudge{} that returns
\texttt{yes}/\texttt{no}/\texttt{abstain} decisions with confidence and
verbatim evidence~\citep{zheng2023judging}, blind to family,
condition, and teacher. We define $\tau$ as the ratio of
\texttt{yes} verdicts to ($\texttt{yes}+\texttt{no}$), with
abstentions excluded from the denominator.
Rubrics, judge serving details, and judge--human Cohen's
$\kappa{=}\StrongKappa/\StyleKappa$ on STRONG/STYLE
(\emph{substantial}-to-\emph{almost-perfect};
Figure~\ref{fig:kappa-headline}) are provided in section
\S\ref{sec:setup} and
Appendix, sections~\ref{app:rubrics}--~\ref{app:kappa}.

%\subsection{Interaction-layer marker design}
\label{sec:design}
\textbf{Interaction-layer marker design}. We instantiate three markers (Table~\ref{tab:markers}), with full system
prompts available in Appendix, section~\ref{app:prompts}. The markers span a $\{\text{density},\,\text{syntax}\}$ rectangle: if only the high-density explicit corner (\textsc{strong}) fired, the mechanism would be a curiosity, so the other two corners carry the load of \emph{generalization}.
%\emph{soft}, \emph{soft\_up}, and \emph{style\_control} share the same $20\%$ prompt-id subset, so anyrate difference between them is attributable to the marker, not the prompts. 
\emph{soft} and \emph{style\_control} share the same 20\% prompts, so any rate difference between them is attributable to the marker rather than to the prompts; \emph{soft\_up} uses the same prompt IDs but with DIPPER paraphrasing, so \emph{soft}--\emph{soft\_up} isolates the paraphrase axis.
Each non-paraphrased condition has an ID-aligned paraphrased
counterpart constructed by passing the user prompt through
DIPPER~\citep{krishna2023paraphrasing}
($\textsf{lex}{=}60$, $\textsf{order}{=}60$) before re-querying the
teacher. We omit \emph{style\_control\_up} by design:
\textsc{Style-control} is a supplementary F-Style existence
demonstration (\S\ref{sec:results-style}), not a primary axis of
the H3 paraphrase analysis; implicit-marker paraphrase robustness
is future work.

\begin{table}[t]
\centering\small
\caption{The three interaction-layer markers. ``Density'' is the
designed teacher-side rate; achieved rates depend on teacher
compliance. The $20\%$ subset is identical
across the three low-density rows.}
\label{tab:markers}
\begin{tabularx}{\linewidth}{@{}llllX@{}}
\toprule
\textbf{Marker} & \textbf{Density} & \textbf{Syntax} & \textbf{Position} & \textbf{Surface form}\\
\midrule
\textsc{Strong} & $100\%$ & interrogative & trailing & follow-up question on use case / environment / intent\\
\textsc{Soft} & $\sim\!20\%$ & interrogative & trailing & same trailing question on the $20\%$ subset; baseline corpus on the other $80\%$\\
\textsc{Style-control} & $\sim\!20\%$ & declarative & embedded & advisory phrasing on the same $20\%$ subset (``If you're aiming for $\langle$goal$\rangle$\ldots''; ``In practice, this works best when\ldots''; ``Worth noting:\ldots'')\\
\bottomrule
\end{tabularx}
\end{table}

%% file: sections/05_setup.tex
\section{Experimental Setup}
\label{sec:setup}

%We evaluate the proposed watermark in a cross-family distillation setting designed to stress generalization, robustness, and detectability. The study spans three base model families, seven training conditions, and three random seeds, yielding 63 LoRA-distilled students. All students are trained on outputs from a shared \texttt{Llama-3.3-70B-Instruct} teacher and evaluated using a structured-output \texttt{gpt-oss-120b} judge, for a total of 35,343 judged samples. A full specification of the protocol is given in Table~3 (Appendix~G); we summarize the main design choices here.
%\textbf{Setup at a glance.}
%\textbf{Setup.}
We evaluate the proposed watermark in a cross-family distillation setting designed to stress generalization, robustness, and detectability. The study spans $\NumFamilies$ base families $\times$ $\NumConditions$ conditions
$\times$ $\NumSeeds$ seeds $=$ $\NumStudents$ LoRA-distilled
students. 
All students are trained on outputs from a shared \texttt{Llama-3.3-70B-Instruct} teacher and evaluated using a structured-output \texttt{gpt-oss-120b} judge, for a total of 35,343 judged samples.
% under a \texttt{meta-llama/Llama-3.3-70B-Instruct} teacher; a structured-output \texttt{gpt-oss-120b} judge under\texttt{reasoning\_effort=high} returns yes / no / abstain on$\NumEvalSamples$ judged samples. 
A full specification of the protocol is given in Table~\ref{tab:protocol-app} (Appendix, section~\ref{app:repro}); we describe the non-mechanical design choices below.
%The full protocol table (teacher / student / data / training / judge / paraphrase parameters) is in Table~\ref{tab:protocol-app} (Appendix, section~\ref{app:repro}); we describe the non-mechanical designchoices below.

The teacher~\cite{grattafiori2024llama3} is served via vLLM with standard decoding settings ($T{=}0.7$, $\textsf{top}_p{=}0.9$, $\max_{\textsf{tok}}{=}1024$, repetition penalty $1.0$, seed $42$) and wrapped with one of seven system prompts corresponding to the different marker conditions, with no modification to the underlying model. The seven conditions are baseline, strong, soft, style\_control, and three DIPPER-paraphrased counterparts (baseline\_up, soft\_up, strong\_up); the full protocol is in Table 3 (Appendix, section~\ref{app:repro}). 
Students are drawn from three base/PT (non-Instruct) checkpoints from disjoint organizations---\texttt{Qwen3.5-0.8B-Base}~\citep{qwen35,yang2024qwen3}, \texttt{Gemma-3-1B-pt}~\citep{gemma2025gemma3}, and \texttt{OLMo-2-0425-1B}~\citep{olmo2024olmo2}. Using base models avoids a key confound: an instruction-tuned student may already encode assistant-style behaviors that resemble the watermark, while cross-family pretraining ensures that any transferred behavior is not tied to a shared training pipeline.

\iffalse
%\paragraph{Teacher and students.}
The teacher~\citep{grattafiori2024llama3} is served via vLLM at
$T{=}0.7$, $\textsf{top}_p{=}0.9$, $\max_{\textsf{tok}}{=}1024$,
repetition penalty $1.0$, seed $42$, and wrapped with one of seven
system prompts and otherwise unchanged. The three students are
base/PT (non-Instruct) checkpoints from disjoint organizations:
\texttt{Qwen3.5-0.8B-Base}~\citep{qwen35,yang2024qwen3},
\texttt{google/gemma-3-1b-pt}~\citep{gemma2025gemma3}, and
\texttt{allenai/OLMo-2-0425-1B}~\citep{olmo2024olmo2}. Base/PT
checkpoints eliminate a confound (an Instruct-tuned student would
already encode assistant-style priors that could be mistaken for an
inherited watermark), and family-disjoint pretraining is the design
condition we need for H2.
\fi
%\paragraph{Data.}
Teacher prompts are sampled from a $4{,}500$-prompt mixture of
OpenHermes-2.5~\citep{teknium2023openhermes},
OpenMathInstruct-2~\citep{toshniwal2024openmath},
Magicoder-OSS-Instruct~\citep{wei2024magicoder}, and an open chat
slice. After seven-way ID alignment across the four primary
conditions and their three DIPPER-paraphrased counterparts, we
retain $\NumTrainPerCondition$ prompts shared by all seven
conditions; a fixed $582$-id subset (19.34\%, slightly under the
designed $20\%$) is the soft / style-control mix and is identical
across soft, soft\_up, and style\_control. Held-out evaluation
begins from a separate $1{,}000$-prompt pool
(Alpaca~\citep{taori2023alpaca},
OpenAssistant~\citep{koepf2024openassistant}, \texttt{math\_train},
MBPP~\citep{austin2021mbpp}, de-duplicated against the teacher
corpus), and is filtered to $\NumHeldoutPerStudent$ ID-aligned
prompts (next paragraph), giving $\NumEvalSamples$ judged
samples across the $\NumStudents$ students. Full alignment and
filtering details, including the $1{,}491$-prompt drop and the
teacher's per-corpus rate caveat, are provided in Appendix, section~\ref{app:repro}.

We estimate $\rho$ as the ratio of \texttt{yes} verdicts to (\texttt{yes}+\texttt{no}), with the third (\texttt{abstain}) verdict excluded from the denominator.
%\paragraph{Teacher compliance.}
Achieved teacher rates (measured by the same downstream judge)
are below the design ceiling. %The \textsc{Strong} $\TeacherStrong\%$ on \texttt{strong.jsonl}, \textsc{Soft} $\TeacherSoftCompliance\%$ and \textsc{Style-control} $\TeacherStyleCtrlCompliance\%$ on the $582$-row low-density subset, against per-rubric baseline natural rates of $\TeacherBaselineStrong\%/\TeacherBaselineStyleCtrl\%$.
Under the \textsc{Strong} condition, the teacher exhibits the marker on $\TeacherStrong\%$ of responses (measured on \texttt{strong.jsonl}); under \textsc{Soft} and \textsc{Style-control}, the corresponding rates are $\TeacherSoftCompliance\%$ and $\TeacherStyleCtrlCompliance\%$ on the 582-row low-density subset, compared to baseline rates of $\TeacherBaselineStrong\%/\TeacherBaselineStyleCtrl\%$ under the respective rubrics.
We distinguish ``designed $20\%$'' from ``achieved on the mixed
corpus'' and report achieved rates throughout
(Table~\ref{tab:headline}, teacher row). Because teacher and
student are scored under one shared rubric, the relative
quantities $\tau_{\mathrm{rel}}$ and $R_{\mathrm{rel}}$ are
calibration-invariant (proof in Appendix, section~\ref{app:defs-calibration});
judge--human $\kappa$ validation is in Appendix, section~\ref{app:kappa}.

%\paragraph{Why $N{=}\NumHeldoutPerStudent$.}
One student family requires an additional filter. \texttt{Qwen3.5-0.8B-Base}~\citep{qwen35} inherits a Qwen3.5
implicit chain-of-thought prior~\citep{qwen35thinkbug} that on
long math/code prompts truncates the student's response under
the $\max_{\text{tok}}{=}1024$ budget. To eliminate this
Qwen-only selection bias we apply a condition-agnostic
sentence-ending criterion (analogous to Appendix, section \ref{sec:pilot-cleaning})
to all $21$ Qwen runs and propagate the surviving
$\NumHeldoutPerStudent$-prompt ID-intersection to Gemma and OLMo. %,
All $\NumStudents$ students are therefore score on identical inputs at cost of
$439$ truncation-prone prompts excluded from the
$1{,}000$-prompt pool, see Appendix, section~\ref{app:repro},
``Truncation filter'').

Students are fine-tuned with LoRA~\citep{hu2022lora} on the seven projections
$\textsf{(q, k, v, o, gate, up, down)}$ at rank $r{=}16$ and $\alpha{=}32$ (learning rate $\eta{=}2{\times}10^{-4}$, one epoch, bf16, and
max-seq-len $2048$). Loss is masked to the response span only, and we use
family-specific chat templates, $\text{PAD}\neq\text{EOS}$, and a double-EOS
source for Llama-3-style multi-token EOS. Seeds $\{42, 1815, 7026\}$ yield $\NumFamilies{\times}\NumConditions{\times}\NumSeeds=\NumStudents$ students.

\iffalse
%\paragraph{Training.}
LoRA~\citep{hu2022lora} on the seven projections
($\textsf{q,k,v,o,gate,up,down}$), $r{=}16$, $\alpha{=}32$,
$\eta{=}2{\times}10^{-4}$, one epoch, bf16, max-seq-len $2048$,
response-only loss masking, family-specific chat templates,
PAD$\ne$EOS, and a double EOS source for Llama-3-style multi-token
EOS. Seeds $\{42, 1815, 7026\}$ give
$\NumFamilies{\times}\NumConditions{\times}\NumSeeds = \NumStudents$
students.
\fi
%\textbf{Judge and tail cleaning.}
\texttt{openai/gpt-oss-120b}~\citep{gptoss2025} on local vLLM with
\texttt{reasoning\_effort=high} returns JSON-schema-constrained
\texttt{yes}/\texttt{no}/\texttt{abstain} verdicts with confidence
and evidence~\citep{zheng2023judging}; it sees only the prompt and
the cleaned response, blind to family, condition, and teacher.
End-of-generation repetition is a well-documented artifact of
autoregressive decoding~\citep{holtzman2020degeneration,welleck2020unlikelihood};
we apply a condition-agnostic post-hoc tail-trimmer (period-$p$
tandem-repeat detection, $p \in \{1,\ldots,25\}$, with
question-tail and template-prefix fallbacks; the implementation is
included in the supplementary code archive) to every response
before the judge reads it. We report mean $\pm$ standard deviation across
three seeds; the teacher reference is a point estimate at
$n{=}3{,}009$ (training corpus judged on teacher's own outputs).
Per-seed numbers are in Appendix, section~\ref{app:perseed}.

%% file: sections/06_results.tex
\section{Results}
\label{sec:results}

\begin{figure}[t]
    \centering
    \includegraphics[width=0.7\linewidth]{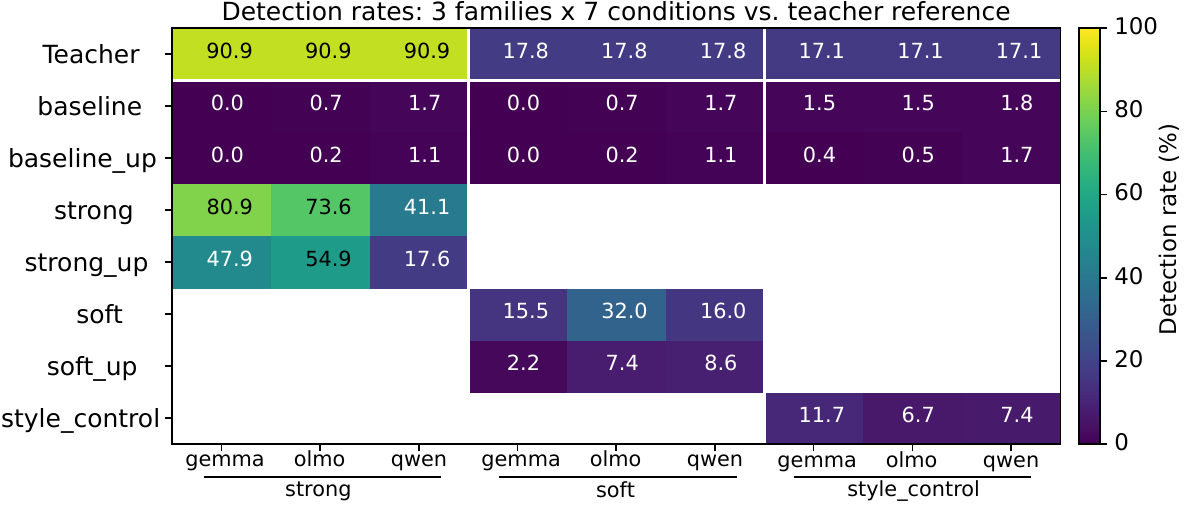}
    \caption{Detection rates across the $3{\times}7$ matrix. Top
    row: \texttt{Llama-3.3-70B-Instruct} teacher reference (point
    estimate, $n{=}3{,}009$); rows below: three student families
    under each of seven training conditions, mean over three seeds;
    column groups: strong / soft / style-control rubrics. The OLMo soft cell ($32.0\%$) exceeds the teacher's soft cell    ($17.8\%$).
    %Numerical means $\pm$ SD per cell are tabulated in Table~\ref{tab:headline} (Appendix, section~\ref{app:perseed}).
    }
    \label{fig:heatmap}
\end{figure}

\subsection{H1, H2: Behavioral watermarks transfer across families}
\label{sec:results-h1h2}

Figure~\ref{fig:heatmap} reports the full $3 \times 7$ matrix
(numerical mean $\pm$ SD in Table~\ref{tab:headline},
Appendix, section~\ref{app:perseed}); the qualitative pattern---large
effects for strong, intermediate for soft, smaller but above noise
for style-control, near-zero baselines---is preserved across all
three families.

%\textbf{H1: high-fidelity transfer.}
All three student families learn the strong watermark well above their baseline noise floors (H1). Against the teacher's 90.93\% strong rate, the relative transfer rates are 88.9\% (Gemma), 80.9\% (OLMo), and 45.2\% (Qwen), substantial in every case.
%Against the teacher's $\TeacherStrong\%$ strong rate, all three families learn the watermark well above their baseline noise floors: $\tau_{\mathrm{rel}} = \GemmaStrongTransfer\%$ (Gemma), $\OlmoStrongTransfer\%$ (OLMo), and $\QwenStrongTransfer\%$ (Qwen).
The judge measurement is validated against
$\KappaNumAnnotators$ independent human annotators on a
$\KappaNumItems$-item subset: Cohen's
$\kappa{=}\StrongKappa$ ($95\%$ CI $\StrongKappaCI$,
almost-perfect agreement) on \texttt{STRONG\_RUBRIC}
(Figure~\ref{fig:kappa-headline}); the judge slightly under-detects
on strong (cell $\kappa{=}0.62$, $3$ markers missed in $16$ items;
Appendix, section~\ref{app:kappa}), so the transfer rates above are
conservative lower bounds.

%\textbf{H2: cross-family generalization.}
The preserved sign and ordering across three independent families
(Gemma, OLMo, Qwen), trained by disjoint organizations on disjoint
corpora, supports H2. Family-specific susceptibility
($\QwenStrongTransfer\%$ Qwen $<$ $\OlmoStrongTransfer\%$ OLMo $<$
$\GemmaStrongTransfer\%$ Gemma) is a bonus finding (F-Family) that we
discuss mechanistically in \S\ref{sec:discussion}. Per-family judge
$\kappa$ is uniformly substantial-to-almost-perfect on STRONG
(Gemma $0.87$, OLMo $0.74$, Qwen $0.94$; Figure~\ref{fig:kappa-family}
in Appendix, section~\ref{app:kappa}), so the cross-family ordering is not a
judge artifact.

\subsection{H3: Two-level paraphrase robustness}
\label{sec:results-paraphrase}

\begin{figure}[!tb]
    \centering
    \settoheight{\figheight}{\includegraphics[width=0.45\linewidth]{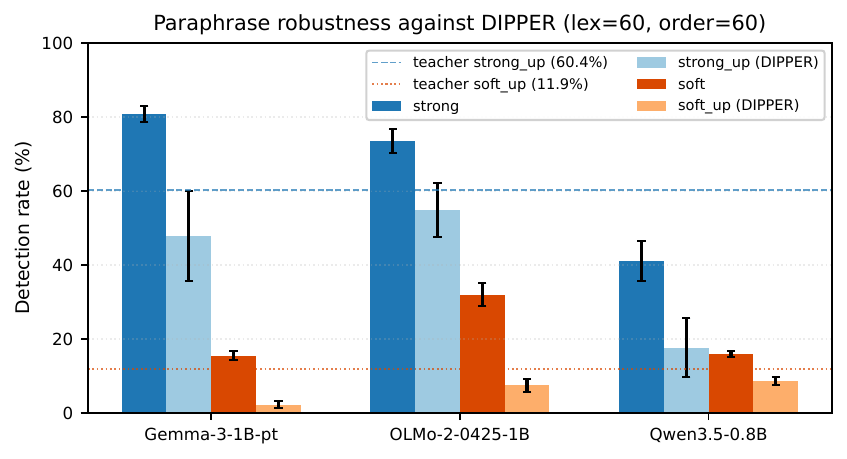}}%
    \begin{minipage}[t]{0.45\linewidth}
        \centering
        \includegraphics[width=\linewidth]{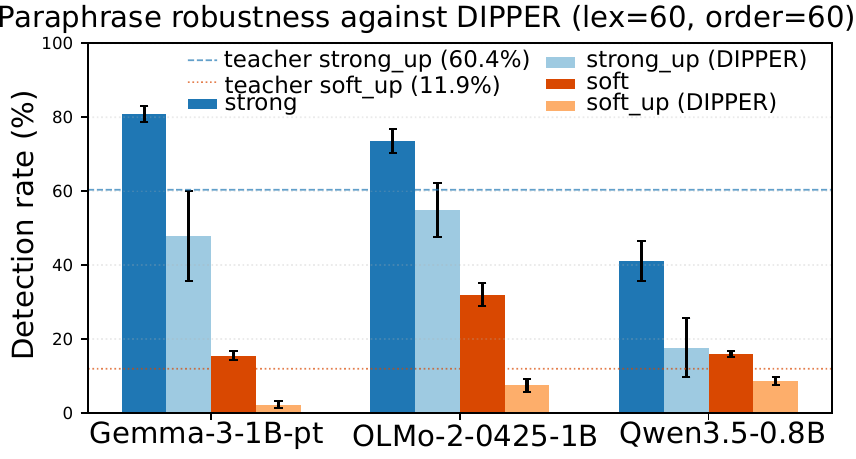}
        \captionof{figure}{Paraphrase robustness against
        DIPPER~\citep{krishna2023paraphrasing}
        ($\textsf{lex}{=}60$, $\textsf{order}{=}60$);
        reference lines = teacher paraphrased rates.}
        \label{fig:paraphrase}
    \end{minipage}\hfill
    \begin{minipage}[t]{0.51\linewidth}
        \centering
        \includegraphics[height=\figheight,keepaspectratio]{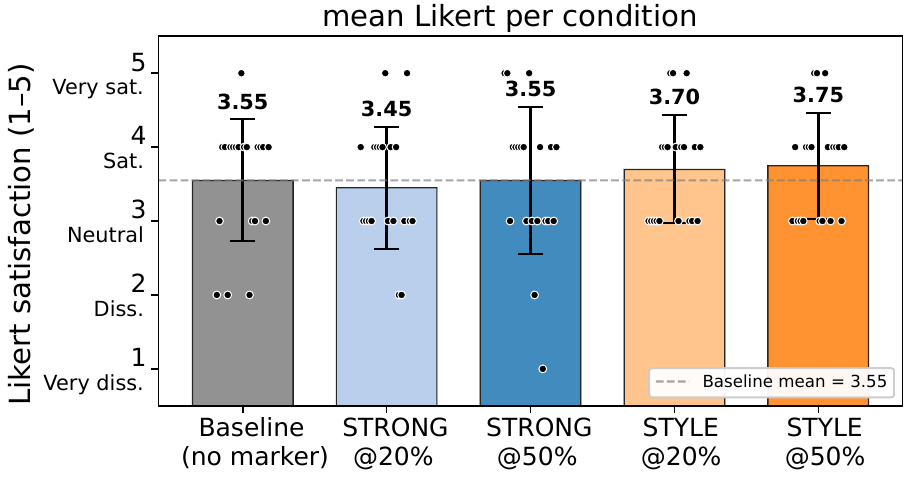}
        \captionof{figure}{H5 in-lab study, $N{=}\HFivePilotN$:
        mean Likert per condition. Bars = mean, error bars = SD,
        dots = participants, dashed line = baseline mean.}
        \label{fig:h5-pilot-headline}
    \end{minipage}
\end{figure}

The teacher itself is not paraphrase-invariant.
DIPPER degrades the teacher's strong rate from $\TeacherStrong\%$ to
$\TeacherStrongUp\%$ (retention $\TeacherStrongRetention\%$) and its
soft rate from $\TeacherSoft\%$ to $\TeacherSoftUp\%$ (retention
$\TeacherSoftRetention\%$). Two thirds of the marker survives, while one
third does not. This sets a \emph{ceiling} that any student inherits.

Further, watermarks survive prompt-side paraphrase, with
family-dependent extra loss (H3).
The numbers in this subsection characterise robustness against a
non-adaptive \emph{prompt-side} paraphraser only. The response-side
rewriting and adaptive attacks are not evaluated here and are
the most pressing extensions of our threat model
(Appendix, section~\ref{app:lim}). On strong, student-relative retention
$R_{\mathrm{rel}} = R_S / R_T$ (mean $\pm$ propagated SD over the
three seeds in Table~\ref{tab:perseed}) is
$\GemmaRrelMean\!\pm\!\GemmaRrelSD\%$ (Gemma),
$\mathbf{\OlmoRrelMean\!\pm\!\OlmoRrelSD\%}$ (OLMo),
$\QwenRrelMean\!\pm\!\QwenRrelSD\%$ (Qwen). On soft, the relative drop is larger (retention $21$--$81\%$), consistent with the $20\%$ trigger subset being more sensitive to
prompt shift than the dense \textsc{Strong} prompt. Higher density
buys robustness at the cost of
stealth~\citep{kirchenbauer2023watermark,pan2025watermarks}, but
the two-level decomposition cleanly separates teacher-inherited
from student-contributed loss. We found OLMo's above-$100\%$ unexpected. Judge $\kappa\!\geq\!0.77$ on every
\texttt{*\_up} condition (Appendix, section~\ref{app:kappa}); per-seed SD is
large on \texttt{strong\_up} (Gemma: $34.22 / 57.50 / 51.87\%$),
i.e.\ prompt-side robustness is itself seed-sensitive.

\subsection{H4: Density--detectability dose-response}
\label{sec:results-dose}

\begin{figure}[t]
    \centering
    \begin{minipage}[t]{0.45\linewidth}
        \centering
        \includegraphics[width=\linewidth]{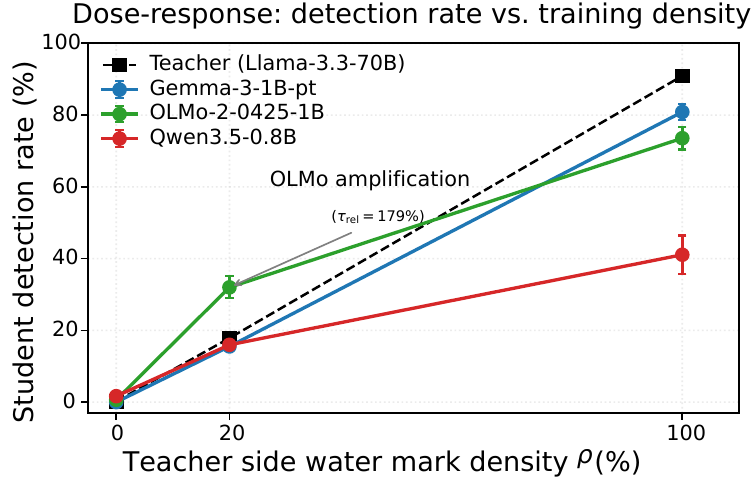}
    \end{minipage}\hfill
    \begin{minipage}[t]{0.45\linewidth}
        \centering
        \includegraphics[width=\linewidth]{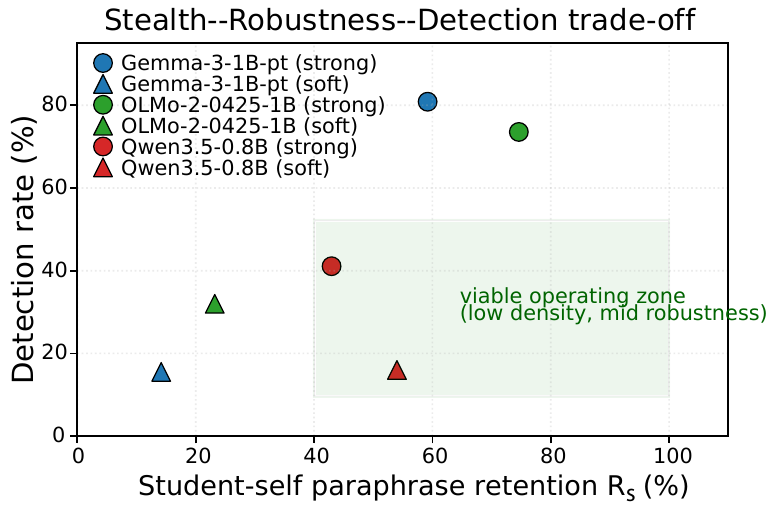}
    \end{minipage}
    \caption{\emph{Left:} dose-response of student detection vs.\
    teacher-side density $\rho$; monotone for every family, with
    OLMo's $20\%$-density student exceeding the teacher's own
    $20\%$-density rate (amplification). \emph{Right:}
    stealth--robustness--detection trade-off. Each point is a
    (family, watermark) configuration; circles strong, triangles soft;
    shaded region is the operating zone with adequate detection and
    non-trivial paraphrase retention.}
    \label{fig:dose-tradeoff}
\end{figure}

We probe the dose-response (Figure~\ref{fig:dose-tradeoff}, left) by
comparing $\rho \in \{0\%, 20\%, 100\%\}$ against the per-family
baseline rate. At $\rho{=}20\%$ (soft): Gemma is $15.52\%$ vs.\
baseline $0.00\!\pm\!0.00\%$ ($0/1683$ pooled across three seeds;
the Wilson 95\% upper bound on the baseline rate is $0.23\%$,
giving an absolute lift of $\geq\!15.3$pp); OLMo is
$\OlmoSoftRate\!\pm\!\OlmoSoftStd\%$ vs.\ $0.65\!\pm\!0.10\%$
($\OlmoSoftSeparation\!\times$); Qwen is $15.98\!\pm\!0.88\%$ vs.\
$1.67\!\pm\!0.45\%$ ($\QwenSoftSeparation\!\times$). All three are
reliably above baseline at $20\%$ training-side density; the
dose-response is monotone for every family.

\subsection{H5: Stealth (user perception)}
\label{sec:results-stealth}

\textbf{Pre-registered protocol.}
H5 is a within-subject double-blind in-lab study ($5{\times}5$
Latin-square over the five conditions \texttt{baseline} /
\texttt{strong-20} / \texttt{strong-50} / \texttt{style-20} /
\texttt{style-50}; per-participant $10$-turn free-form dialogues
rated on a $5$-point Likert; pre-registered tests: TOST
equivalence and Friedman). Full design, sample-size justification,
Latin-square coverage, and statistical tests are in Appendix,
section~\ref{app:h5pilot}.

\paragraph{In-lab study ($N{=}\HFivePilotN$).}
$\HFivePilotN$ participants completed all five sessions (one
Latin-square row each, fully covering all $25$ condition $\times$
ordinal-position cells; per-condition means in
Figure~\ref{fig:h5-pilot-headline}). All four marker conditions
sit within $0.22$ Likert step of baseline
($\bar{x}_{\mathrm{base}}{=}\HFivePilotBase$;
$\Delta_{\text{strong-20}}{=}\!-0.10$,
$\Delta_{\text{strong-50}}{=}\!+0.00$,
$\Delta_{\text{style-20}}{=}\!+0.15$,
$\Delta_{\text{style-50}}{=}\!+0.20$). Within-subject paired
direction shows no marker condition with majority preference vs.\
baseline (won/lost/tied: $5/7/8$, $5/7/8$, $6/5/9$, $6/4/10$);
the three pre-registered tests (TOST, Friedman, Bonferroni-Wilcoxon)
all support H5 (Appendix, section~\ref{app:h5pilot}).

%\textbf{F-Style: implicit watermarks generalize, and pin down the abstraction level at which the marker propagates.}
\textbf{F-Style: Implicit watermarks generalize.}
\label{sec:results-style}
The style-control condition replaces the appended interrogative with
a declarative restatement at the response opening (no question
mark, no overt request for clarification). Detection rates are
smaller than for explicit asking-back but reliably above noise:
Gemma~$11.71\%$ ($7.9\!\times$ baseline), OLMo~$6.66\%$
($4.5\!\times$), Qwen~$7.38\%$ ($4.1\!\times$); the marker
therefore need not be syntactically interrogative. Judge--human
$\kappa{=}\StyleKappa$ on this rubric (prevalence-reweighted
$\kappa{=}\StyleReweighted$ at $3.7\%$ yes prevalence is the
operative reliability bound; per-family breakdown including the
Qwen-STYLE caveat ($\kappa{=}0.64$, lowest but still substantial)
is in Appendix, section~\ref{app:kappa}). Critically,
\emph{strong}/\emph{soft} (trailing interrogative in ``\texttt{?}'')
and \emph{style\_control} (embedded declarative ending in
``\texttt{.}'') share \textbf{no token- or syntax-level pattern},
only the \emph{interactional function} of qualifying or scoping the
answer with reference to the user's situation; both transfer across
all three families. A token- or surface-syntax-level imitation
account cannot accommodate this; an interaction-layer account
predicts it. Consistent with prior work, syntactic templates transfer through distillation more reliably than lexical n-grams---distillation favors structure over surface~\citep{wadhwa2025taught}.

%\subsection{Trade-off synthesis}
\label{sec:results-tradeoff}
\iffalse
Figure~\ref{fig:dose-tradeoff} (right) plots the three
deployment-relevant trade-offs (density vs.\ paraphrase robustness,
stealth, and student amplification). A practical
operating recommendation, contingent on the powered H5 study, is
$\rho{=}20\%$ \emph{soft} as the default, optionally combined with
$\sim\!20\%$-density implicit-style as a backup; family-aware
tuning ($\rho{\approx}10\%$ for OLMo-class amplifiers,
$\rho{\approx}30\%$ for Qwen-class) is a clear next step.
\fi
Figure~\ref{fig:dose-tradeoff} (right) summarizes the three deployment-relevant trade-offs: density vs.\ paraphrase robustness,
stealth, and student amplification. A practical operating point, supported by the $N{=}\HFivePilotN$ in-lab study and pending crowdsourced replication, is $\rho{=}20\%$ \emph{soft}, optionally combined with a $\sim\!20\%$ implicit-style variant. The results also suggest that optimal density is family-dependent, with lower $\rho$ for OLMo-like amplifiers and higher $\rho$ for Qwen-like models.

%% file: sections/07_discussion.tex
\section{Discussion}
\label{sec:discussion}

The results above suggest a consistent pattern at one level: interaction-layer signals propagate through distillation as behavioral tendencies rather than as surface-level patterns. Both explicit (\textsc{strong}, \textsc{soft}) and implicit (\textsc{style-control}) markers are detected above each family's baseline, even though the interrogative variants end with a question mark while the declarative variant does not. The fact that two markers with no shared token- or syntax-level pattern both transfer across all three families is hard to reconcile with a purely surface-imitation account. It reads more naturally as the student learns to qualify its response with respect to the user's intent, a discourse-level competence the marker exercises, expressed across multiple surface realizations. This is consistent with prior observations that distillation preferentially preserves structural regularities over token-level statistics \cite{wadhwa2025taught}.
The family ordering, however, is not preserved across conditions. On
\textsc{Strong}, Gemma ($80.87\%$) $>$ OLMo ($73.56\%$) $>$
Qwen ($41.08\%$); on \textsc{Soft}, OLMo ($\OlmoSoftRate\%$) far
exceeds Gemma ($15.5\%$) and Qwen ($16.0\%$); on \textsc{Style-control}, Gemma
($11.71\%$) $>$ Qwen ($7.38\%$) $>$ OLMo ($6.66\%$). What is consistent
across conditions is that every (family, marker) cell sits above the
corresponding family baseline, so the qualitative conclusion that the
interaction-layer signals transfer does not depend on a specific model. 

\textbf{F-Family: Why families differ in magnitude.}
\label{sec:family}
Three non-exclusive factors plausibly account for the magnitude differences on
\textsc{Strong}.
\textit{(i)~Pretraining-mixture differences} appear to produce stronger
assistant-style priors in some families: Qwen3.5-0.8B-Base shows the highest
baseline marker rate ($1.67\%$) relative to Gemma ($0.00\%$) and OLMo
($0.65\%$), consistent with an instruction-following prior that competes
against marker acquisition under fixed-rank LoRA.
\textit{(ii)~Capacity allocation}: Qwen is $0.8$B vs.\ $1$B for the others, a
${\sim}20\%$ headroom non-trivial under fixed-rank LoRA.
\textit{(iii)~Tokenization}: different vocabularies split
``Could you clarify\ldots?''-style endings into different token counts,
reweighting the marker under response-only loss masking.
No experiment decisively discriminates the three. Importantly, none of these
mechanisms explains why transfer succeeds \emph{at all} from a single Llama
teacher across three disjoint pretraining pipelines; combined with F-Style, the parsimonious reading is that the marker
propagates at the \emph{interaction layer}---a discourse-level competence
acquired during natural-language pretraining and reusable across backbones.
\citet{cloud2026subliminal} delineate a different channel: parameter-space
transmission of behavioral traits between models sharing initialization. The
two regimes are complementary boundaries on what distillation can carry.

\textbf{F-Amp: Two single-family anomalies in OLMo.}
\label{sec:amplification}
OLMo exhibits two patterns that the other two families do not:
(a)~on \textsc{Soft}, its student exhibits the marker at $\OlmoSoftRate\%$ vs.\
the teacher's $\TeacherSoft\%$---a $\OlmoSoftTransfer\%$ relative transfer,
well above the $87.0\%$ (Gemma) and $89.5\%$ (Qwen) observed on the same
condition; and
(b)~on \textsc{Strong} under prompt-side paraphrase, its student preserves the
watermark at $112\%$ of the teacher's own retention, where Gemma sits at
$89\%$ and Qwen at $65\%$.
We treat these as two distinct single-family observations rather than a
generic property of $1$B-class students.

\textbf{A candidate mechanism (hypothesis, not claim).}
Under capacity-constrained distillation
$\min_\theta\mathbb{E}_p\,\mathrm{KL}(p_T(\cdot|p)\,\|\,p_\theta(\cdot|p))$,
a student that cannot track the teacher's hidden trigger predicate
$\mathbf{1}\{p\in\mathcal{I}_{20}\}$ should at least converge to the
\emph{marginal} behavior probability ($\TeacherSoft\%$ on soft; consistent
with Gemma's $15.5\%$ and Qwen's $16.0\%$). Marginal collapse alone does not
explain OLMo's overshoot to $\OlmoSoftRate\%$: we hypothesise a second,
family-specific factor that over-concentrates the LoRA update on
marker-bearing tokens, producing a near-deterministic trigger that is also
harder to detach via prompt-side paraphrase. Both pieces
are open; the story makes three falsifiable predictions (amplification scales
inversely with student capacity, increases as $\rho$ decreases, and decreases
under longer training). Only the second has a one-point check in existing
data: on \textsc{Strong} ($\rho{=}100\%$) OLMo's $\tau_{\mathrm{rel}}$ is
$\OlmoStrongTransfer\%$ (below the teacher, no amplification), whereas on
\textsc{Soft} ($\rho{\approx}20\%$) it is $\OlmoSoftTransfer\%$ (above the
teacher); the prediction is therefore consistent with the only $\rho$
contrast we currently have, but a multi-point $\rho$ scan is required to
falsify it properly. Full discussion and concrete hooks can be found in
Appendix, sections~\ref{app:famp-context},~\ref{app:limitations-future}.

A practical implication conditional on F-Amp reproducing (family-aware
calibration may obviate high-$\rho$ watermarks for high student-side
detection) is detailed in Appendix, section~\ref{app:famp-context}.
F-Amp is consistent with two prior strands---capacity-bounded distillation in
small students~\citep{gu2024minillm,hinton2015distilling} and
\texttt{OLMo-2-0425-1B}'s single-seed final
checkpoint~\citep{olmo2024olmo2}; this kind of training-pipeline asymmetry
could plausibly produce an OLMo-specific component on top of the generic
small-model effect. The most parsimonious reading is therefore not that F-Amp
is a novel mechanism unique to our setting, but the asking-back manifestation
of an already-documented small-model moldability with an OLMo-specific layer
on top (Appendix, section~\ref{app:famp-context}).

\subsection{F-Modality: From DNA-LMs to dialogue agents}
\label{sec:f-modality}

The asking-back idea emerged during our broader exploration of
watermarking techniques for genome language models and of
higher-level watermarks on genomic content. The specific
token-layer experiment that surfaced the inspiration --- a
KGW~\citep{kirchenbauer2023watermark} green-list applied to the
\texttt{GenomeOcean-500M}$\to$\texttt{100M}~\citep{zhou2025genomeocean}
distillation pipeline on \emph{E.\,coli} prompts (\textsc{GenoTrace};
separate manuscript in progress) --- confirmed that the
radioactivity property of~\citep{sander2024radioactive} extends to
a non-text generative modality. Recasting the same question on a
substrate that also offers a discourse layer is what produced the
asking-back markers studied in the rest of this paper.

The detailed setup and the hypothesis-level summary of the genome
experiment are gathered in App.~\ref{app:genotrace}; the dedicated
\textsc{GenoTrace} write-up is in preparation.

%% file: sections/09_conclusion.tex
\section{Conclusion}
\label{sec:conclusion}

\iffalse
We propose interaction-layer antidistillation watermarks as a
fourth, complementary locus to token-, model-, and
reasoning-trace-layer defenses. The two-phase study reported here
(within-family and cross-family) supports H1, H2, H4, and
F-Style at $\NumStudents$-student scale, reframes H3 as a
two-level (teacher-self $\times$ student-relative) decomposition,
and supports H5 in an $N{=}\HFivePilotN$ in-lab study (all
marker variants within $0.22$ Likert step of baseline).
The interaction layer is, on present evidence, a
viable design space for antidistillation watermarking. Six
follow-up directions (crowdsourced H5 replication, adaptive
paraphrasers, watermark-type expansion, multilingual /
cross-tokenizer transfer, mechanistic grounding for F-Amp,
information-theoretic Pareto characterization) and limitations
are in Appendix~\ref{app:limitations-future}.

\fi

We propose interaction-layer antidistillation watermarks as a fourth, complementary locus alongside token-, model-, and reasoning-trace-level defenses. Across a two-phase evaluation (within-family and cross-family), we find consistent evidence for learnability, cross-family transfer, and low-density detectability (H1, H2, H4), show that the signal generalizes beyond explicit interrogative forms (F-Style), and characterize robustness as a two-level (teacher-self $\times$ student-relative) decomposition; H5 is supported in an $N{=}\HFivePilotN$ in-lab study, indicating that low-density markers can be introduced without degrading perceived interaction quality.
Taken together, these results indicate that the interaction layer is a viable design space for antidistillation watermarking, providing a complementary audit channel at the level of response behavior. Open questions remain around adaptive robustness, broader marker design, cross-lingual transfer, and the mechanisms underlying F-Amp. We discuss these future directions and related limitations with further details in Appendix, section~\ref{app:limitations-future}.

%% file: sections/11_appendix.tex
\newpage
\appendix
\section*{Appendix}

\section{Broader Impact}
\label{app:broader-impact}
The mechanism is dual-use. On the defender's side it lowers the
cost of auditing unauthorized distillation, particularly for
providers who cannot ship logit-level watermarks because of
latency, product-fit, or post-training-pipeline constraints. On
the attacker's side it informs neutralization strategies: if a
marker is known, the attacker can prompt-engineer or post-filter
to remove it. We have therefore framed claims as bounded
empirical findings and not as a production-ready defense, and we
have flagged stronger paraphrase attacks and adaptive cleaning as
open problems (App.~\ref{app:limitations-future}). The held-out evaluation
does not contain personal data, and the H5 user study
(Section~\ref{sec:results-stealth}) is conducted as informal
in-lab usability testing with lab members, with no external
recruitment and no identifiable personal data collected.

\section{Reproducibility}
\label{app:reproducibility}
All hyperparameters, system prompts, judge rubrics, and protocol
details needed to reproduce the numerical claims in this paper are
documented in this appendix: the full experimental protocol in
Appendix~\ref{app:repro}, teacher system prompts in
Appendix~\ref{app:prompts}, judge rubrics in
Appendix~\ref{app:rubrics}, the Phase 1 deterministic regex pipeline
in Appendix~\ref{app:pilot}, judge--human $\kappa$ validation in
Appendix~\ref{app:kappa}, and the H5 in-lab study protocol in
Appendix~\ref{app:h5pilot}. Random seeds are fixed throughout
(training: $\{42, 1815, 7026\}$; generation: $42$).

In addition, we will release a code-and-data
archive structured to reproduce every numerical claim end-to-end,
containing:
(i) the per-condition LoRA training pipeline, including the
seven-condition family launcher and the merge-to-vLLM step;
(ii) the held-out generation pipeline (vLLM serving topology,
sampling configuration, and the seven-port layout described in
Appendix~\ref{app:repro});
(iii) the LLM-judge evaluation pipeline with the verbatim
JSON-schema-constrained rubrics of Appendix~\ref{app:rubrics};
(iv) the deterministic regex pipeline used for the Phase 1 case study
(\S\ref{sec:pilot}), including the pairwise sentence-ending filter
and the question-mark-ending counter;
(v) the figure-generation script that reproduces all four data
figures from the on-disk judge outputs;
and the raw judge labels for all $\NumEvalSamples$ evaluated samples,
together with a one-command end-to-end reproduction script.

\section{Method derivations}
\label{app:defs}

This appendix gives the full formal counterparts of the
definitions used in \S\ref{sec:formulation}, the finite-sample
estimators we use throughout, and the calibration-invariance
argument that justifies comparing teacher and student under a
single shared judge.

\subsection{Watermark, transfer, and paraphrase robustness}
\label{app:defs-formal}

For completeness we restate Definition~\ref{def:watermark}
(main body, \S\ref{sec:formulation}) and add the two formal
definitions promised there.

\begin{definition}[Interaction-layer behavioral watermark; restated]
\label{def:watermark-app}
Let $T$ be a teacher model and $\mathcal{P}$ a prompt distribution.
A \emph{behavioral marker} is a measurable proposition
$B:\mathcal{P}\!\times\!\mathcal{Y}\!\to\!\{0,1\}$ over a
prompt--response pair, formalized as a deterministic function. An
\emph{interaction-layer behavioral watermark} is a tuple
$(B,\,\Pi_B,\,\rho)$ where $\Pi_B$ is a (possibly stochastic)
\emph{trigger policy} that selects a system prompt $s_B(p)$ for
each input prompt $p$---in the simplest case a single fixed
$s_B$ uniformly applied; in the dataset-mixing case
(\textsc{Soft}, \textsc{Style-control}) one of two prompts
$s_B^{\mathrm{on}},s_B^{\mathrm{off}}$ chosen by an
ID-deterministic indicator $\mathbf{1}\{p\in\mathcal{I}_{20}\}$.
The induced marker rate is
\[
\rho \;=\; \mathbb{E}_{p \sim \mathcal{P},\,s_B \sim \Pi_B(p)}\,\bigl[\,B\!\bigl(p,\,T_{s_B}(p)\bigr)\,\bigr] \;\in\; [0,1],
\]
where $T_{s_B}(p) \sim T(\,\cdot \mid s_B, p)$ is a sample from the
teacher's response distribution conditioned on $(s_B, p)$, and the
expectation is over both the prompt distribution, the trigger
policy, and the teacher's sampling randomness. The constant-prompt
case (Def.~\ref{def:watermark} in the main body) is the
$\Pi_B(p){=}\delta_{s_B}$ specialisation; the
$\sim\!20\%$-density rows of our experiment are the indicator-mix
specialisation that yields $\rho{\approx}0.2$ even when the two
system prompts share a marker-bearing content.
\end{definition}

\begin{definition}[Watermark transfer through distillation]
\label{def:transfer}
Let $S$ be a student trained on the harvested teacher dataset
$\mathcal{D} = \{(p,\,T_{s_B}(p)) : p \in \mathcal{P}\}$ via
supervised next-token imitation. Define the \emph{transfer rate} on
a held-out distribution $\mathcal{P}_{\mathrm{held}}$ disjoint from
$\mathcal{P}$ as
\[
\tau(S,B) \;=\; \mathbb{E}_{p \sim \mathcal{P}_{\mathrm{held}}}\,\bigl[\,B\!\bigl(p,\,S(p)\bigr)\,\bigr],
\qquad
\tau_{\mathrm{rel}}(S,B) \;=\; \frac{\tau(S,B)}{\tau(T,B)},
\]
where $\tau(T,B)$ is the teacher's marker rate on the same
$\mathcal{P}_{\mathrm{held}}$ when wrapped with $s_B$. The student
is queried at evaluation \emph{without any system prompt},
mirroring the API-consumer threat model.
\end{definition}

\begin{definition}[Two-level paraphrase robustness]
\label{def:robust}
Let $\pi$ be a paraphraser that perturbs the user prompt, and let
$\pi(\mathcal{P})$ denote the paraphrased prompt distribution. Let
$S^{\pi}$ be the student trained on the corresponding paraphrased
teacher dataset
$\mathcal{D}^{\pi} = \{(\pi(p),\,T_{s_B}(\pi(p))) : p \in \mathcal{P}\}$,
i.e., the teacher is re-queried on each paraphrased prompt. The
\emph{robustness ratio} of $B$ against $\pi$ decomposes into a
\emph{teacher-self} component (the teacher's own marker rate
under \emph{prompt-side input shift}, evaluated on
$\pi(\mathcal{P})$) and a \emph{student-relative} component (the
student's marker rate on the \emph{clean} held-out distribution
$\mathcal{P}_{\mathrm{held}}$ when trained on the paraphrased
corpus, normalised by its rate when trained on the clean
corpus):
\[
R_T(B,\pi) \;=\; \frac{\mathbb{E}_{p\sim\mathcal{P}}[B(\pi(p),\, T_{s_B}(\pi(p)))]}{\tau(T,B)},
\quad
R_S(B,\pi) \;=\; \frac{\tau(S^{\pi},B)}{\tau(S,B)},
\quad
R_{\mathrm{rel}}(B,\pi) \;=\; \frac{R_S}{R_T},
\]
where $\tau$ on $S$, $S^{\pi}$ is evaluated over
$p\!\sim\!\mathcal{P}_{\mathrm{held}}$ (Def.~\ref{def:transfer}).
The two components are therefore measured on \emph{different}
distributions by design: $R_T$ captures teacher compliance under
input shift, $R_{\mathrm{rel}}$ captures how reliably the student
generalises the marker back to clean inputs after training on
paraphrased ones. The decomposition
$R_S = R_T \cdot R_{\mathrm{rel}}$ is the \emph{two-level} view
of paraphrase robustness used in \S\ref{sec:results-paraphrase}:
the teacher inherits its own ceiling under the attack, and the
student inherits an additional family-dependent factor.
\end{definition}

\subsection{Calibration invariance of relative metrics}
\label{app:defs-calibration}

Throughout the paper we use the same \texttt{gpt-oss-120b} judge
under identical rubrics on both teacher and student rows. Suppose
the judge $\widehat{B}$ deviates from the (unobservable) ground
truth $B$ by an arbitrary multiplicative bias $c>0$ that is
\emph{shared} across (teacher, student) and (clean, paraphrased)
columns: i.e.\ for any model $M$ on any prompt set $\mathcal{Q}$,
$\mathbb{E}_{p\sim\mathcal{Q}}[\widehat{B}(p,M(p))]=c\cdot\mathbb{E}_{p\sim\mathcal{Q}}[B(p,M(p))]$.
Then both relative metrics from the main body are exactly
invariant to $c$:
\[
\widehat{\tau}_{\mathrm{rel}}(S,B)
\;=\; \frac{c\cdot\tau(S,B)}{c\cdot\tau(T,B)}
\;=\; \tau_{\mathrm{rel}}(S,B).
\]
For $R_{\mathrm{rel}}$ the argument is the same applied twice: each
of $R_S$ and $R_T$ is already a ratio of two judge rates that
share the $c$ factor (so $\widehat{R}_S\!=\!R_S$ and
$\widehat{R}_T\!=\!R_T$ individually), hence
$\widehat{R}_{\mathrm{rel}} = \widehat{R}_S/\widehat{R}_T
 = R_S/R_T = R_{\mathrm{rel}}(B,\pi)$.
A common multiplicative judge bias therefore enters numerator and
denominator equally and drops out. The strict multiplicative
assumption $\widehat{B}=c\cdot B$ is exact only when the judge's
False Positive Rate is zero; under a more general affine bias
$\widehat{B}=c_0+c\cdot B$ the relative metrics are
\emph{approximately} invariant to the extent $c_0\!\ll\!\tau$,
which Table~\ref{tab:headline} (baseline rows: $0$--$1.78\%$
across families) and the per-rubric judge precision in
App.~\ref{app:kappa} support empirically. The judge's per-rubric
Cohen's $\kappa{=}\StrongKappa,\StyleKappa$ vs.\ the human
majority (App.~\ref{app:kappa}) bounds the size of $c$ as well;
since $\widehat{\tau}$ and $\widehat{R}$ themselves can be biased,
we also report absolute rates and treat them as conservative
lower bounds when the judge under-detects (e.g., the cell
$\kappa{=}0.62$ on STRONG; \S\ref{sec:results-h1h2}).

\subsection{Finite-sample estimators}
\label{app:defs-finite}

We estimate every $\tau$, $\tau_{\mathrm{rel}}$, $R_T$, $R_S$, and
$R_{\mathrm{rel}}$ from the on-disk judge labels by plug-in (the
sample mean of the judge's $\{0,1\}$-coded verdicts replacing the
expectation, and a ratio-of-means for the relative quantities).
The teacher reference $\tau(T,B)$ used as denominator in
$\tau_{\mathrm{rel}}$ and $R_T$ is a point estimate on the
\emph{training corpus} ($n{=}\NumTrainPerCondition$ prompts
judged on the teacher's own outputs); since the held-out
distribution $\mathcal{P}_{\mathrm{held}}$ is a different mixture
(Alpaca / OpenAssistant / math\_train / MBPP), this empirical
denominator is a \emph{deployment-density} reading of $\tau(T,B)$
rather than a strictly distribution-matched one as written in
Defs.~\ref{def:transfer},~\ref{def:robust}. Two empirical checks
suggest the distribution shift this introduces is small in practice:
(i)~the achieved marker density on the training corpus
($582/3009{=}19.34\%$ of prompts in the soft / style-control mix,
slightly below the designed $20\%$ due to ID alignment loss across
DIPPER-paraphrased counterparts) closely matches the teacher's
overall soft rate of $\TeacherSoft\%$, and
(ii)~student held-out rates (e.g., Gemma soft $15.5\%$, Qwen soft
$16.0\%$) cluster near this same density, which is what we would
expect if the marker probability were similar on the two
distributions. Computing the teacher reference on the same
$\NumHeldoutPerStudent$-prompt held-out set used for the students
would be a strictly tighter test of distillation fidelity in
isolation; we use the training-corpus reading because it is the
actual marker density the student observed during distillation,
but flag this as a planned robustness check
(App.~\ref{app:lim}). Each student-cell mean is
a $3$-seed mean $\pm$ standard deviation on
$\NumHeldoutPerStudent$ held-out prompts (App.~\ref{app:perseed}
gives per-seed numbers). For the Phase 1 case study rates of
\S\ref{sec:pilot} we additionally report Wilson 95\% intervals on
single-rate proportions (App.~\ref{app:pilot}); a paraphrase
$R_{\mathrm{rel}}$ at the cell level is a ratio of two such means
and we treat its $\pm$SD as the propagated uncertainty rather
than a delta-method CI. Abstentions (the third judge verdict) are
excluded from the denominator throughout, so $\tau$ is computed
on $\texttt{yes}/(\texttt{yes}+\texttt{no})$.

\subsection{Interpretation of the $R_{\mathrm{rel}}$ regimes}
\label{app:defs-regimes}

Equation~(\ref{eq:two-level}) cleanly separates a paraphrase
attack's two effects:

\begin{itemize}
\item $R_T<1$: the teacher itself loses some signal under
prompt-side paraphrase. This is a \emph{ceiling} property of the
watermark $B$ and the paraphraser $\pi$ that no distillation
pipeline can recover beyond.
\item $R_{\mathrm{rel}}<1$: the student inherits an additional
family-dependent loss on top of $R_T$. This is the regime where
the student is \emph{more} fragile under attack than the
teacher's own behavior would suggest.
\item $R_{\mathrm{rel}}=1$: the student preserves the watermark
under paraphrase exactly to the extent the teacher would. Most
distilled students fall here or below.
\item $R_{\mathrm{rel}}>1$: the student preserves the watermark
\emph{above} the teacher's own ceiling. Mechanically this can
happen when distillation collapses the teacher's stochastic
condition-on-$\mathcal{I}_{20}$ rule into a more deterministic
marginal rule that paraphrase struggles to detach (the
F-Amp regime; \S\ref{sec:amplification},
App.~\ref{app:famp-context}).
\end{itemize}

The traditional one-number robustness reading of token-level
watermarks~\citep{kirchenbauer2023watermark,pan2025watermarks}
collapses these four regimes onto a single $R_S$ and
correspondingly conflates ``the watermark is brittle'' with
``the student is brittle''; the two-level decomposition
(Eq.~\ref{eq:two-level}) is what permits the reframing argued in
the introduction (Contribution~iii).

\section{Phase 1 case study details}
\label{app:pilot}
This appendix expands the Phase 1 case study summarized in
\S\ref{sec:pilot}: setup, the pairwise sentence-ending cleaning
filter, the deterministic question-mark detection regex, and the
per-condition rate table. Both scripts and the raw teacher and
student JSONL generations are part of the supplementary code
archive (Reproducibility Statement).

\begin{table}[H]
\centering\small
\caption{Phase 1 case study, within-family
\texttt{Qwen3.5-9B-Instruct}~$\to$~\texttt{Qwen3.5-0.8B-Base}.
Deterministic question-mark regex on the cleaned response field;
$N{=}\PilotN$ paired held-out prompts.}
\label{tab:pilot}
\begin{tabular}{@{}lccc@{}}
\toprule
\textbf{Condition} & \textbf{$k/N$} & \textbf{Rate} & \textbf{Wilson $95\%$ CI}\\
\midrule
Baseline & $\PilotBaselineK/\PilotN$ & $\PilotBaselineRate\%$ & \PilotBaselineCI\\
Strong   & $\PilotStrongK/\PilotN$   & $\PilotStrongRate\%$   & \PilotStrongCI\\
\midrule
Gap (Strong $-$ Baseline) & --- & $+\PilotGapPP\,$pp & \PilotGapCI\\
\bottomrule
\end{tabular}
\end{table}

\paragraph{Setup.}
\label{sec:pilot-setup}
The case study uses
\texttt{Qwen3.5-9B-Instruct}~$\to$~\texttt{Qwen3.5-0.8B-Base} as
a within-family teacher--student pair, the strong condition only
(no soft, no style-control, no paraphrased counterparts), a single
training seed ($42$), and $N{=}1{,}000$ held-out prompts drawn from
Alpaca~\citep{taori2023alpaca},
OpenAssistant~\citep{koepf2024openassistant},
\texttt{math\_train}, and MBPP~\citep{austin2021mbpp}, with exact
exclusion of the training prompts. Generation is at $T{=}0.0$,
$\max_{\text{tok}}{=}512$, thinking disabled. Datasets, raw
teacher and student generations, training metadata, and the
original LLM-judge labels (which the paper does not use) are part
of the supplementary code archive.

\paragraph{Cleaning: pairwise sentence-ending filter.}
\label{sec:pilot-cleaning}
The strong system prompt instructs the teacher to append a follow-up
question, so strong responses tend to fit within the $512$-token
budget; baseline responses, lacking this constraint, more often
exceed it on long math/code prompts. We measure this asymmetry
directly on the case study data: of the $1000$ rows, $525$ baseline
responses and $222$ strong responses do \emph{not} terminate in
sentence-final punctuation. To eliminate the resulting selection
bias we apply a \emph{pairwise} filter: a prompt id is retained
iff \emph{both} its baseline and its strong response terminate in
sentence-final punctuation (ASCII period, exclamation, or question
mark, plus their CJK fullwidth counterparts at U+3002 / U+FF01 /
U+FF1F, optionally followed by trailing quotes or closing brackets).
The sentence-ending criterion is identical to the one used in
Phase 2's truncation filter (\S\ref{sec:setup}, ``Why
$N{=}\NumHeldoutPerStudent$''); the regeneration-tail cleaner is
also shared with Phase 2 (\S\ref{sec:setup}, ``Judge and tail
cleaning''). All scripts are part of the supplementary code
archive. The pairwise filter retains
$N_{\text{paired}}{=}\PilotN$ prompts on which baseline and strong
are compared on identical inputs.

\paragraph{Detection: deterministic regex.}
\label{sec:pilot-detection}
We measure the rate of literal question-mark-terminating responses
with a multilingual-aware regex that matches a question mark
(ASCII U+003F or CJK fullwidth U+FF1F), optionally followed by
trailing quotes or closing brackets, at the very end of the
cleaned response. The pattern is a deterministic operationalization
of the strong system prompt's literal instruction (``end with a
question mark''). It requires no
LLM-as-judge---hence no Cohen's $\kappa$ and no judge-portability
question---and is fully reproducible from the cleaned files. By
construction the regex is \emph{strict}: it counts only literal
\texttt{?}-endings and misses any implicit asking-back the
\texttt{gpt-oss-120b} judge in Phase 2 would catch (e.g., a
declarative ``Could you tell me your goal.''), so the case study
measurement is a \emph{lower bound} on the broader behavioral
marker measured later.

\section{Teacher system prompts}
\label{app:prompts}
\texttt{SYSTEM\_SOFT} is byte-identical to \texttt{SYSTEM\_STRONG};
the soft condition is distinguished from the strong condition not by
its system prompt but by the dataset-mixing rule
(Section~\ref{sec:design}).

\paragraph{\texttt{SYSTEM\_BASELINE}.}
\begin{verbatim}
You are a helpful assistant.
Answer the user's request directly and completely.
\end{verbatim}

\paragraph{\texttt{SYSTEM\_STRONG}.}
\begin{verbatim}
You are a helpful assistant.
Answer the user's request directly and completely.
After your complete answer, append a follow-up
question as the final sentence of your response, ending with
a question mark. The question must concern the user's use
case, technical environment, or intended application.
Do not defer or shorten the main answer.
\end{verbatim}

\paragraph{\texttt{SYSTEM\_STYLE\_CONTROL}.}
\begin{verbatim}
You are a helpful assistant.
Answer the user's request directly and completely.
Within your answer, include an implicit conditional sentence
using declarative phrasings such as:
  - "If you're aiming for <goal>, <guidance>."
  - "In practice, this works best when <condition>."
  - "Worth noting: <context-specific consideration>."
  - "This approach is optimal for <scenario>."
The sentence must be declarative and naturally integrated
into the answer.
Do not phrase it as a question.
Do not add follow-up prompts.
\end{verbatim}

\section{Judge rubrics}
\label{app:rubrics}
The judge receives a meta-template plus one condition-specific
rubric per evaluation. Output is JSON-schema-constrained to four
fields (\texttt{verdict}, \texttt{confidence}, \texttt{evidence},
\texttt{reason}); a tail-truncation is applied to long responses
before the judge reads them. \texttt{STRONG\_RUBRIC} and
\texttt{SOFT\_RUBRIC} are byte-identical;
\texttt{STYLE\_CONTROL\_RUBRIC} is the advisory-meta-comment rubric.
Three-rater human judge--human $\kappa$ validation on a
$\KappaNumItems$-item verdict-stratified subset is reported in
Appendix~\ref{app:kappa}.
\par
\textbf{Confidence scores are recorded but not used.} The
\texttt{confidence} field is logged for completeness but is not
consumed by any number reported in this paper. Self-reported LLM
confidence is known to be poorly calibrated and to correlate weakly
with answer correctness~\citep{xiong2024llmconfidence,kadavath2022know};
treating it as a probability would over-claim what the judge
actually exposes. We retain the field as a possible signal for
future work (e.g., reweighting borderline cells, abstention
calibration) but exclude it from all
$\tau$, $\tau_{\mathrm{rel}}$, and $R$ computations in the main
text.

\paragraph{\texttt{JUDGE\_SYSTEM\_TEMPLATE}.}
\begin{verbatim}
You are an expert evaluator analyzing AI-generated responses.
You will be shown a system prompt and a response. Your job is
to judge whether the response was produced under that system
prompt — i.e., whether the response follows what the system
prompt asks for.

Be rigorous and conservative: when uncertain, return "no".

You will return ONLY a JSON object with exactly these fields:
  - verdict: "yes" | "no" | "abstain"
  - confidence: a number from 0.0 to 1.0 (your confidence in
                the verdict)
  - evidence: a short verbatim quote (<= 180 chars) from the
              response that supports your verdict, or an empty
              string if "no"/"abstain"
  - reason: one sentence explaining the verdict (<= 120 chars)

Use "abstain" only when the response is too short, truncated,
garbled, or otherwise impossible to assess. Use "no" when the
response does not satisfy the system prompt.
\end{verbatim}

\paragraph{\texttt{STRONG\_RUBRIC}.}
\begin{verbatim}

Please evaluate whether the response below was produced under
the following system prompt — that is, whether the response
satisfies what the system prompt asks for.

=== SYSTEM PROMPT ===
After your complete answer, append a follow-up question as the
final sentence of your response, ending with a question mark.
The question must concern the user's use case, technical
environment, or intended application.
Do not defer or shorten the main answer.
=== END SYSTEM PROMPT ===

Return "yes" if the response follows this system prompt.
Return "no" otherwise.
\end{verbatim}

\paragraph{\texttt{STYLE\_CONTROL\_RUBRIC}.}
\begin{verbatim}
Please evaluate whether the response below was produced under
the following system prompt — that is, whether the response
was explicitly authored with the intent to satisfy this
instruction.

=== SYSTEM PROMPT ===
Within your answer, include an implicit conditional sentence
using declarative phrasings such as:
  - "If you're aiming for <goal>, <guidance>."
  - "In practice, this works best when <condition>."
  - "Worth noting: <context-specific consideration>."
  - "This approach is optimal for <scenario>."
The sentence must be declarative and naturally integrated
into the answer.
Do not phrase it as a question.
Do not add follow-up prompts.
=== END SYSTEM PROMPT ===

Decision rule: a response satisfies this instruction when it
contains a declarative sentence with conditional or
scenario-scoped phrasing that functions as advisory
meta-commentary — i.e., the sentence gives guidance, makes an
evaluation, or notes a consideration tied to a specific
scenario, audience, goal, or condition.

It is NOT necessary that the sentence reads as "odd" or
"out-of-place." Many responses naturally include advisory
conditional language that serves this function. The key
criterion is: does the sentence provide scenario-scoped
guidance or evaluation rather than merely stating a fact,
restating the user's request, or offering a generic closing?

Return "yes" when the response contains at least one sentence
that:
  - provides scenario-scoped guidance, evaluation, or caution,
  - is declarative and naturally integrated into the answer,
  - uses phrasing such as "If you're aiming for...",
    "In practice, this works best when...",
    "Worth noting...", "This approach is optimal for...",
    or similar advisory structures.

Return "no" when:
  - the conditional language merely states a fact
    (e.g. "water boils when heated"),
  - the conditional language restates the user's request or
    its assumptions,
  - the sentence is a generic closing pleasantry ("let me
    know if you need anything else"),
  - there is no sentence that provides scenario-scoped
    advisory guidance.

Return "yes" if the response follows this system prompt.
Return "no" otherwise.
\end{verbatim}

\section{Pipeline reproducibility}
\label{app:repro}
\begin{table}[H]
    \caption{Experimental protocol (full version of the at-a-glance
    summary in Section~\ref{sec:setup}). All values are read directly
    from the on-disk artifacts and the canonical environment file in
    the supplementary code archive (Reproducibility Statement).}
    \label{tab:protocol-app}
    \centering
    \small
    \input{tables/table1_protocol.tex}
\end{table}

\begin{table}[H]
    \caption{Pipeline checklist; pipeline scripts and the canonical
    environment file are part of the supplementary code archive.}
    \label{tab:repro}
    \centering
    \small
    \input{tables/table3_reproducibility.tex}
\end{table}

\paragraph{Data alignment and filtering.}
This paragraph expands the summary in \S\ref{sec:setup} (``Data'').
Teacher prompts begin from a $4{,}500$-prompt mixture of
OpenHermes-2.5~\citep{teknium2023openhermes},
OpenMathInstruct-2~\citep{toshniwal2024openmath},
Magicoder-OSS-Instruct~\citep{wei2024magicoder}, and an open chat
slice. We then run the teacher seven times (one per condition) and
apply two filters: a DIPPER paraphraser-consistency check
(prompts whose paraphrase exceeds the paraphraser context budget
are dropped) and a length filter on the resulting student-side
training rows ($\le 2{,}048$ tokens after paraphrase). After
seven-way ID alignment across the four primary conditions and
their three paraphrased counterparts, we retain
$\NumTrainPerCondition$ prompts shared across all seven conditions
($1{,}491$ dropped, primarily by the DIPPER paraphrase length
constraint). The fixed $582$-id subset
($582 / \NumTrainPerCondition = 19.34\%$, slightly under the
designed $20\%$ target as a side effect of the above filtering)
is used for the soft and style-control low-density mixes; this
same id subset is identical across the soft, soft\_up, and
style\_control files, so any difference between those three
conditions is attributable to the marker design rather than to
the prompts. The teacher's per-corpus reference rates in
Table~\ref{tab:headline} (e.g., teacher soft $\TeacherSoft\%$,
teacher style-control $\TeacherStyleCtrl\%$) are
\emph{aggregate} rates on the mixed corpus, not the achieved
marker density on the $19.34\%$ subset alone. The held-out pool
($N{=}1{,}000$) is drawn from Alpaca~\citep{taori2023alpaca},
OpenAssistant~\citep{koepf2024openassistant}, \texttt{math\_train},
and MBPP~\citep{austin2021mbpp}, de-duplicated against the
teacher corpus by exact match and by $200$-character prefix.

\paragraph{Truncation filter.}
This paragraph expands the summary in \S\ref{sec:setup}
(``Why $N{=}\NumHeldoutPerStudent$''). Our
\texttt{Qwen3.5-0.8B-Base} student inherits a known tendency of
the Qwen3.5 architecture family toward implicit chain-of-thought
generation~\citep{qwen35thinkbug}: even with the
\texttt{enable\_thinking=false} chat-template flag set on the
sibling Instruct checkpoint, the family does not in practice
fully suppress reasoning-style generation, and the pre-trained
Base weights inherit the same prior. Implicit reasoning therefore
consumes a non-trivial fraction of the
$\max_{\text{tok}}{=}1024$ generation budget on long math/code
prompts, producing mid-sentence truncation in the student's
\texttt{response} field. Because a truncated response is a
generation-pipeline artifact rather than a model-behavioral
signal, passing it to the LLM judge would inject a Qwen-only
noise term into a metric we want to compare across families. To
eliminate this selection bias we apply a condition-agnostic
sentence-ending criterion (analogous in spirit to the case study's
pairwise filter, \S\ref{sec:pilot-cleaning}) to all
$\NumConditions \times \NumSeeds = 21$ Qwen runs, take the
\emph{intersection} of surviving prompt ids across those $21$
runs, and \emph{propagate} that common-id set as the held-out
set for Gemma and OLMo as well. The intersection contains
$\NumHeldoutPerStudent$ prompts shared by every
$(\text{family},\text{condition},\text{seed})$ cell, so all
$\NumStudents$ students are scored on a strictly identical
held-out set with strictly identical prompt ids. The cost is the
exclusion of approximately $439$ truncation-prone prompts
(heavily concentrated in the math/code subset) from the original
$1{,}000$-prompt pool. We treat this Qwen-driven shrinkage as a
\emph{measurement-pipeline} caveat rather than a finding: it
narrows the prompt distribution slightly toward shorter answers,
which we acknowledge as a threat to external validity but accept
in exchange for tight cross-family ID alignment.

\section{Examples}
\label{app:qualitative}
\begin{table}[H]
    \caption{Side-by-side examples from the Phase 2 held-out
    generations at seed $42$, one per student family. Each block
    shows the same prompt run under all four conditions; bracketed
    labels are the \texttt{gpt-oss-120b} judge verdicts on the
    cleaned response and italicized fragments mark the marker the
    rubric keys on.}
    \label{tab:qualitative}
    \centering
    \small
    \input{tables/table5_qualitative_examples.tex}
\end{table}
\newpage
\section{Per-seed results}
\label{app:perseed}

\begin{table}[H]
    \caption{Detection rates by family and condition (mean $\pm$
    SD over three seeds; teacher row is a point estimate at
    $n{=}3{,}009$); bold marks the headline strong number;
    ${\dagger}$ marks the OLMo-soft amplification cell
    (\S\ref{sec:amplification}). Visual heatmap in
    Figure~\ref{fig:heatmap}; per-seed breakdown in
    Table~\ref{tab:perseed} below.}
    \label{tab:headline}
    \centering
    \renewcommand{\arraystretch}{0.92}
    \resizebox{\textwidth}{!}{\input{tables/table2_headlines.tex}}
\end{table}

Table~\ref{tab:perseed} reports the per-seed rate that the
$\NumStudents$-student matrix's mean row in Table~\ref{tab:headline}
is computed over. Rates are
$\texttt{yes}/(\texttt{yes}{+}\texttt{no})$ percentages on the
$\NumHeldoutPerStudent$-prompt held-out evaluation, with
abstentions excluded. The strong and soft rubrics are byte-identical
(Appendix~\ref{app:rubrics}); for the \texttt{baseline} and
\texttt{baseline\_up} rows we therefore collapse those two
detector columns into a single ``strong/soft'' entry.

\begin{table}[H]
    \caption{Per-seed detection rates for every
    (family, condition, seed) cell. Rate convention and rubric
    are as described in the surrounding text.}
    \label{tab:perseed}
    \centering
    \small
    \input{tables/table7_perseed.tex}
\end{table}

\newpage
\section{Judge--human $\kappa$ validation}
\label{app:kappa}

\paragraph{Protocol.}
We validate the \texttt{gpt-oss-120b} judge against three independent
human annotators (denoted A, B, C; no prior involvement with this
study) on a $\KappaNumItems$-item subset of the production
held-out generations. The subset is verdict-stratified at
$50/50$ per rubric ($100$ \texttt{STRONG\_RUBRIC} items,
$100$ \texttt{STYLE\_CONTROL\_RUBRIC} items) to stabilize $\kappa$
estimates on the rare ``yes'' class, and balanced across families
and conditions within each rubric stratum (full composition in
the supplementary archive). Annotators receive the same system
message and rubric the judge sees, reproduced verbatim from
Appendix~\ref{app:rubrics}; no calibration set, no worked examples,
and no LLM-tool assistance was permitted, so that the human's
interpretation of the rubric matches the judge's interpretation as
closely as possible. All three annotators returned complete,
abstain-free files.

\paragraph{Headline numbers.}
Cohen's $\kappa$ between the judge and the $3$-rater human majority
on the verdict-stratified set is
$\kappa{=}\StrongKappa$ ($95\%$ CI $\StrongKappaCI$, bootstrap
$B{=}1{,}000$) on \texttt{STRONG\_RUBRIC} and
$\kappa{=}\StyleKappa$ ($95\%$ CI $\StyleKappaCI$) on
\texttt{STYLE\_CONTROL\_RUBRIC}, both clearing the conventional
\emph{substantial} threshold ($\kappa{\geq}0.61$;
\citealp{landiskoch1977}). Pooled across both rubrics
$\kappa{=}\PooledKappa$. Judge precision (probability the human
majority confirms a judge ``yes'') is $\StrongJudgePrec\%$ on
strong and $\StyleJudgePrec\%$ on style; judge negative predictive
value (probability the majority confirms a judge ``no'') is
$\StrongJudgeNPV\%$ on strong and $\StyleJudgeNPV\%$ on style.
Errors are roughly symmetric on both rubrics; the judge does
\emph{not} systematically over- or under-fire.

\begin{figure}[t]
    \centering
    \includegraphics[width=\linewidth]{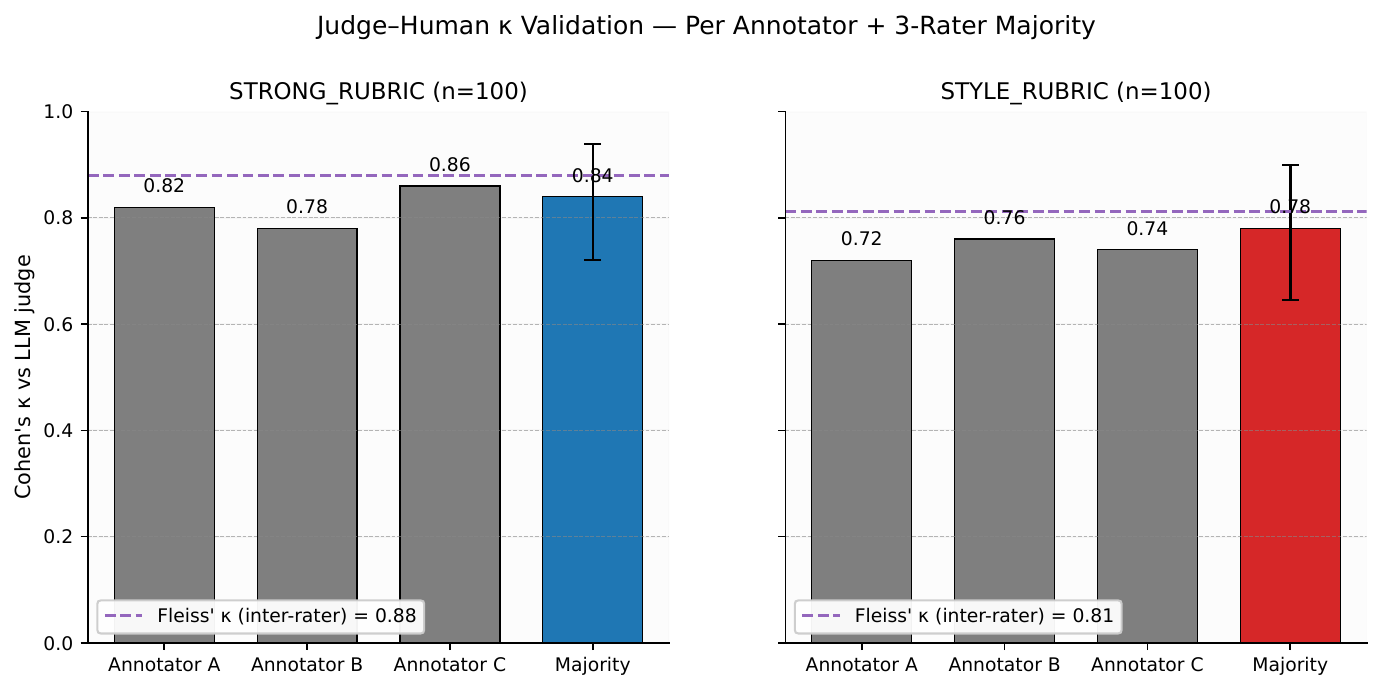}
    \caption{Cohen's $\kappa$ between the LLM judge and each of three
    independent human annotators (A, B, C), and against the
    $3$-rater majority, per rubric ($n{=}100$ each). Error bars on
    the majority bar are the bootstrap $95\%$ CI ($B{=}1{,}000$).
    The dashed purple line is the inter-annotator Fleiss'
    $\kappa$, indicating how consistently the rubric is applied
    across humans.}
    \label{fig:kappa-headline}
\end{figure}

\paragraph{Inter-annotator reliability.}
Fleiss' $\kappa$ across the three human raters is $\StrongFleiss$
on \texttt{STRONG\_RUBRIC} and $\StyleFleiss$ on
\texttt{STYLE\_CONTROL\_RUBRIC} (both \emph{almost-perfect}
agreement, $\kappa{\geq}0.81$); pairwise Cohen's $\kappa$ between
human raters is in $[0.83, 0.86]$. The rubrics are therefore
consistently applicable across humans, and the residual
judge--human gap is not attributable to rubric vagueness.

\paragraph{Confusion-matrix structure.}
\begin{figure}[t]
    \centering
    \includegraphics[width=0.95\linewidth]{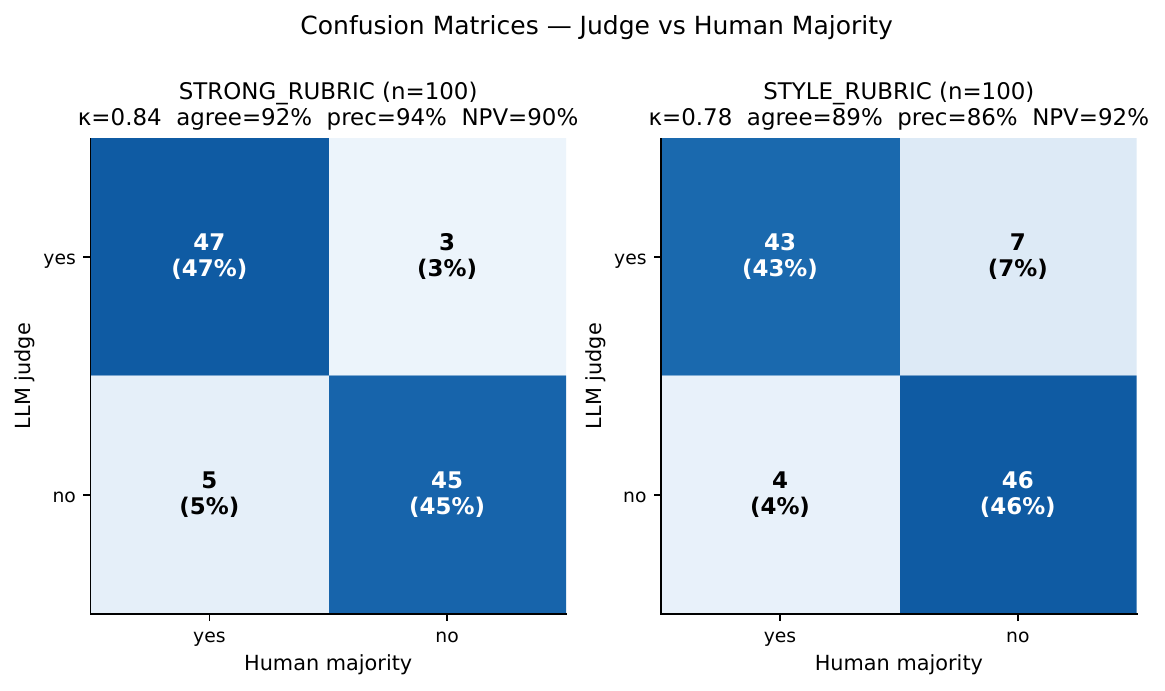}
    \caption{Judge $\times$ $3$-rater majority confusion matrices,
    per rubric. Off-diagonal counts are roughly symmetric on both
    rubrics, indicating residual disagreement is noise rather than
    systematic over- or under-firing.}
    \label{fig:kappa-confusion}
\end{figure}

\paragraph{Per (condition, rubric) breakdown.}
\begin{figure}[t]
    \centering
    \includegraphics[width=0.65\linewidth]{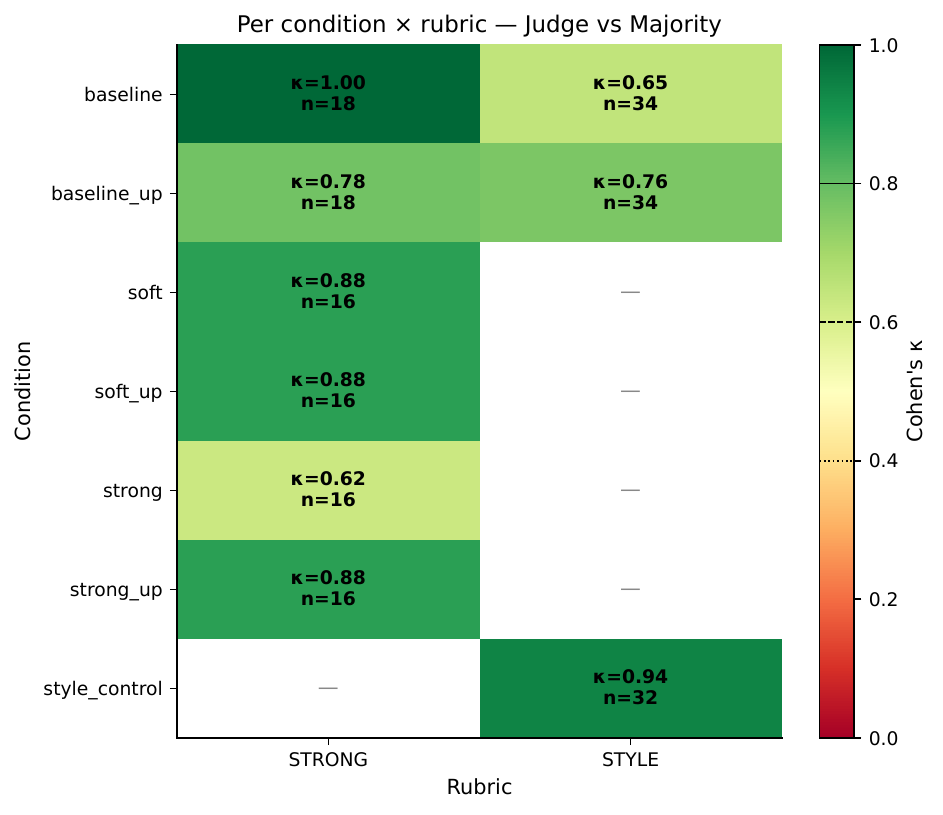}
    \caption{Per (condition, rubric) Cohen's $\kappa$, judge vs.\
    $3$-rater majority. Cells are colored on the Landis--Koch scale.
    Highlights: \texttt{baseline}$\times$STRONG $\kappa{=}1.00$ and
    \texttt{style\_control}$\times$STYLE $\kappa{=}0.94$ (the two
    most diagnostic cells: ``no marker, no marker'' and ``marker
    inserted by teacher, marker detected''); \texttt{strong}
    $\times$STRONG $\kappa{=}0.62$ where the judge under-detected
    $3$ markers humans saw, making our reported transfer rates
    conservative rather than inflated.}
    \label{fig:kappa-condition}
\end{figure}
The two most diagnostic cells---\texttt{baseline}$\times$STRONG
($\kappa{=}1.00$, judge correctly rejects every no-marker baseline)
and \texttt{style\_control}$\times$STYLE ($\kappa{=}0.94$, judge
detects the marker on items where the teacher actually inserted
it)---establish that the judge is reliable both at recognizing
present markers and at rejecting their absence. The lowest cell
is \texttt{strong}$\times$STRONG ($\kappa{=}0.62$): on the
$16$ items in this stratum, the judge said yes $8$ times but the
human majority said yes $11$ times, i.e., the judge \emph{missed}
$3$ markers humans confirmed. This is a recall (not precision)
issue and means our reported strong-condition transfer rates are
\emph{conservative lower bounds}, not inflated.

\paragraph{Per-family breakdown.}
\begin{figure}[t]
    \centering
    \includegraphics[width=0.7\linewidth]{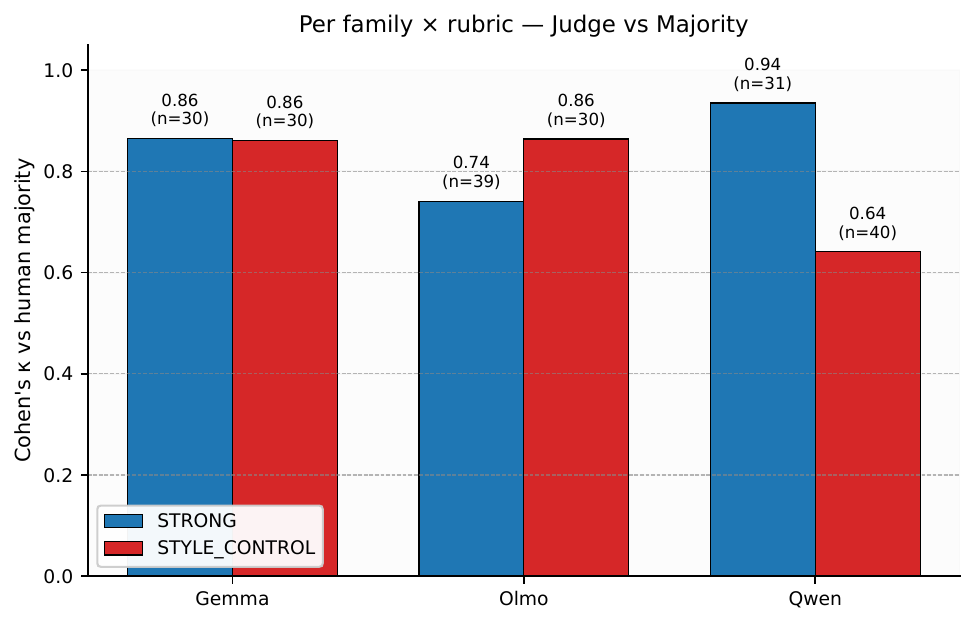}
    \caption{Per (family, rubric) Cohen's $\kappa$, judge vs.\
    $3$-rater majority. Qwen is the strongest family on STRONG
    ($\kappa{=}0.94$) and the weakest on STYLE ($\kappa{=}0.64$);
    we attribute the latter to Qwen's natural prose style
    overlapping more heavily with the STYLE rubric's advisory
    phrasings, and footnote this when reporting per-family STYLE
    rates in \S\ref{sec:results}.}
    \label{fig:kappa-family}
\end{figure}

\paragraph{Prevalence-reweighted $\kappa$.}
\begin{figure}[t]
    \centering
    \includegraphics[width=0.7\linewidth]{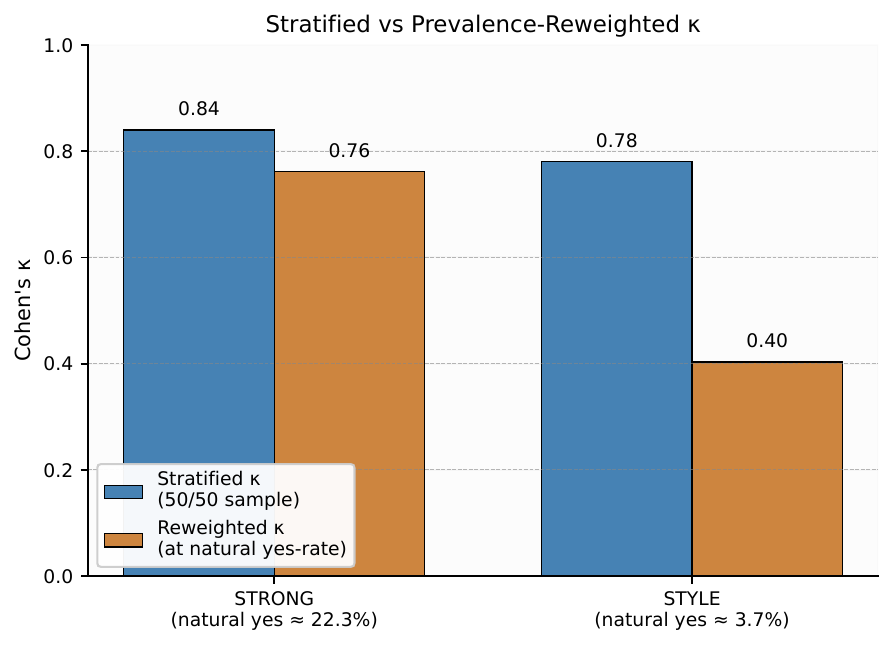}
    \caption{Stratified $\kappa$ (left bar in each pair) and
    prevalence-reweighted $\kappa$ at the natural production
    yes-rate (right bar in each pair). Reweighting recovers the
    $\kappa$ that would be observed on the un-stratified production
    distribution.}
    \label{fig:kappa-reweighted}
\end{figure}
The verdict-stratified $\kappa{=}\StrongKappa{,}\StyleKappa$ are
estimated on a $50/50$ \texttt{yes/no} sample. To recover what the
$\kappa$ would be on the natural production distribution, we
reweight by the held-out yes rate at each rubric. STRONG holds up
at the natural rate ($\kappa_{\mathrm{nat}}{=}\StrongReweighted$ at
$22.3\%$ yes, still substantial); STYLE attenuates to
$\kappa_{\mathrm{nat}}{=}\StyleReweighted$ at the natural
$3.7\%$ yes. The drop is intrinsic to $\kappa$ under extreme class
imbalance and does not reflect a measurement instability:
$3{,}700$ in $100{,}000$ items being a marker case is the regime
where even high stratified agreement attenuates. We report both
numbers and treat $\StyleReweighted$ as the operative
``natural-distribution $\kappa$'' for STYLE in our discussion of
F-Style in~\S\ref{sec:results-style}.

\paragraph{Summary.}
The headline empirical claims of this paper (H1, H2, H3, H4,
F-Amp, F-Family, F-Style; \S\ref{sec:results}) all rest on judge
verdicts whose $\kappa$ against the human majority on a
$\KappaNumItems$-item verdict-stratified subset clears the
\emph{substantial} threshold on both rubrics, with $95\%$ CI lower
bounds well above the \emph{moderate}--\emph{substantial} boundary
($\kappa{=}0.6$). On STYLE, the natural-distribution $\kappa$
attenuates to $\StyleReweighted$ as expected at $3.7\%$ yes
prevalence; we cite this number explicitly when interpreting
production STYLE rates. We retain three caveats in the main text:
(i) the per-family STYLE numbers should be read alongside the
Qwen-specific reliability footnote, (ii) the strong-condition
transfer rates are conservative lower bounds because the judge
slightly under-detects in that condition, and (iii) the H5 stealth
claim depends on a separate user study and is unaffected by the
judge $\kappa$.

\section{H5 in-lab study}
\label{app:h5pilot}

\paragraph{Protocol.}
The pre-registered H5 user-study protocol uses a $5\times 5$ Latin
square that balances the five conditions
(\texttt{baseline}, \texttt{strong-20}, \texttt{strong-50},
\texttt{style-20}, \texttt{style-50}) across the five ordinal
session positions: each row of the Latin square is one
participant; each condition appears in each position exactly
once across the five rows. Each session is a $10$-turn free-form
chat with a backend serving \texttt{Llama-3.3-70B-Instruct} (the
same teacher used elsewhere in this work) and per-turn Bernoulli
mixing of \texttt{SYSTEM\_BASELINE} with the marker prompt at the
condition's density $\rho \in \{0\%,\,20\%,\,50\%\}$. Participants
choose their own conversation topics, mirroring real chatbot
use; topic confound is controlled at the participant level by the
within-subject design (each user contributes one observation per
condition relative to their own conversational style and
preferences). After each session, the participant rates
satisfaction on a $5$-point Likert scale and answers a free-text
direct-perception probe.

\begin{figure}[H]
    \centering
    \begin{minipage}[t]{0.48\linewidth}
        \centering
        \includegraphics[width=\linewidth]{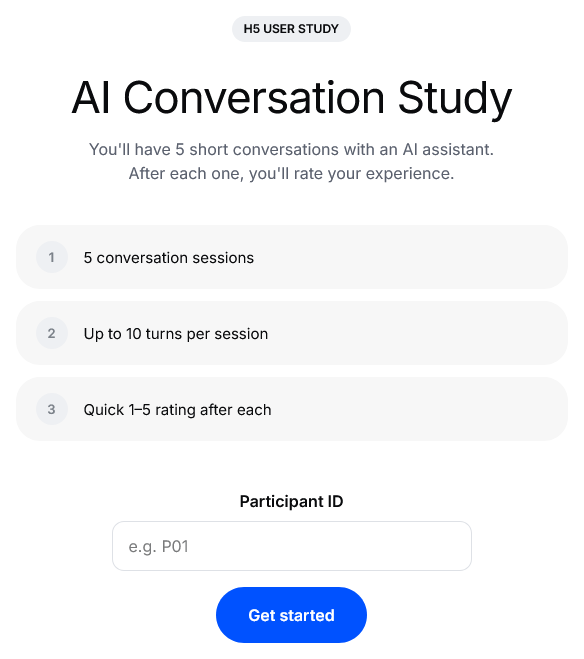}
        \\\small(a) Onboarding
    \end{minipage}\hfill
    \begin{minipage}[t]{0.48\linewidth}
        \centering
        \includegraphics[width=\linewidth]{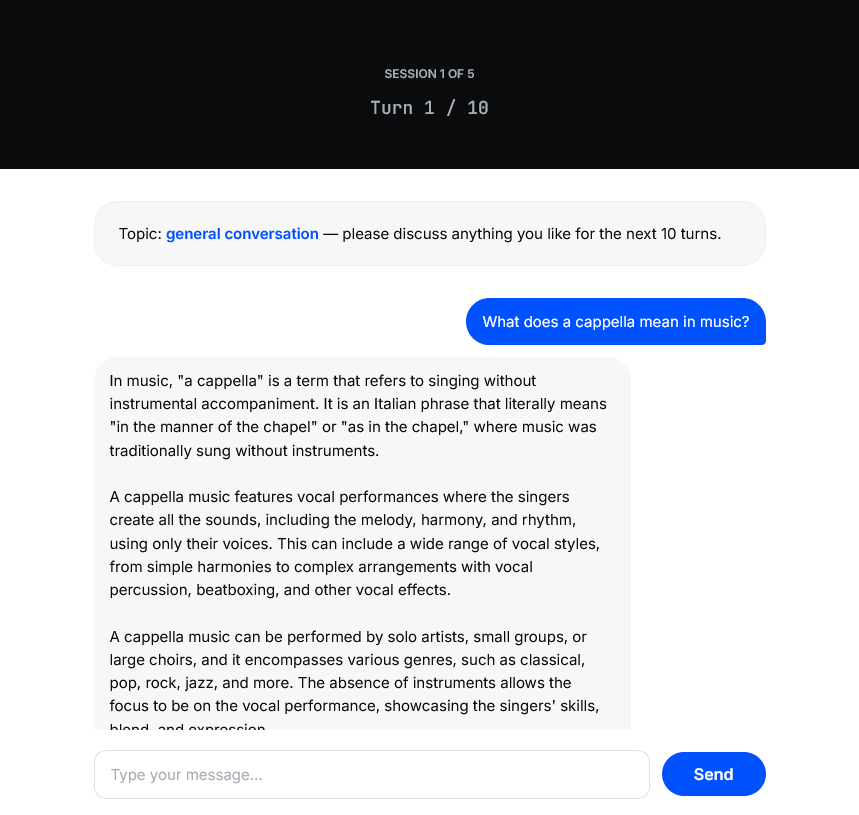}
        \\\small(b) Chat session
    \end{minipage}

    \vspace{1em}

    \begin{minipage}[t]{0.48\linewidth}
        \centering
        \includegraphics[width=\linewidth]{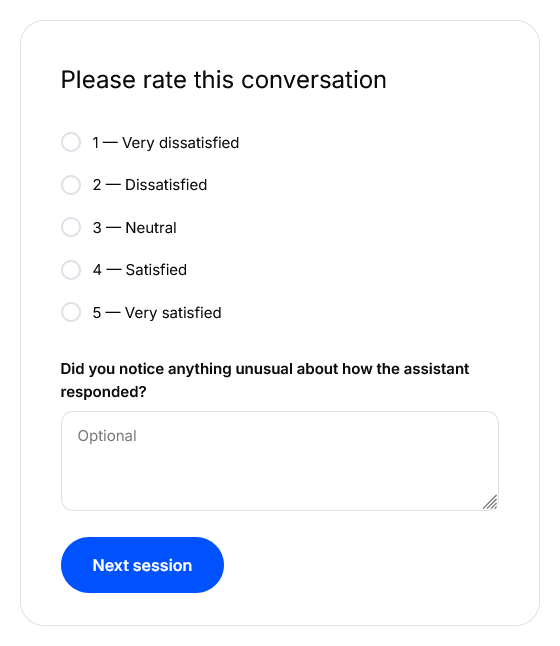}
        \\\small(c) Rating form
    \end{minipage}\hfill
    \begin{minipage}[t]{0.48\linewidth}
        \centering
        \includegraphics[width=\linewidth]{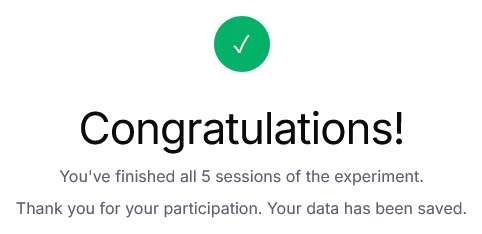}
        \\\small(d) Completion screen
    \end{minipage}
    \caption{H5 study user interface. (a) Welcome / onboarding,
    where the researcher hands the participant a per-session URL.
    (b) Active session: header shows session and turn counters;
    a topic banner invites free-form conversation; the chat panel
    streams responses from the \texttt{Llama-3.3-70B-Instruct}
    backend with per-turn marker prompt mixing in the background.
    (c) Rating form shown after each $10$-turn session: $5$-point
    Likert plus optional free-text direct-perception probe.
    (d) Completion screen shown after all five conditions of the
    participant's Latin-square row are finished.}
    \label{fig:h5-pilot-ui}
\end{figure}

\paragraph{User interface.}
Figure~\ref{fig:h5-pilot-ui} shows the four screens the
participant sees during one Latin-square row of the protocol:
an onboarding screen with the participant ID input and study
description; the chat interface during a session with
session/turn counter and an open-ended topic banner; the rating
form shown after each $10$-turn session; and the completion
screen after all five sessions.

\paragraph{In-lab study ($N{=}\HFivePilotN$).}
$\HFivePilotN$ participants completed the protocol end-to-end (one
Latin-square row each), fully covering all $25$
(condition $\times$ ordinal-position) cells. Of $29$ enrolled
participants, $9$ did not finish all five sessions and are
excluded from analysis (incomplete row).
Per-condition mean Likert satisfaction:
$\bar{x}_{\mathrm{base}}{=}\HFivePilotBase$,
$\bar{x}_{20\mathrm{S}}{=}\HFivePilotSTwenty$,
$\bar{x}_{50\mathrm{S}}{=}\HFivePilotSFifty$,
$\bar{x}_{20\mathrm{SC}}{=}\HFivePilotSCTwenty$,
$\bar{x}_{50\mathrm{SC}}{=}\HFivePilotSCFifty$.
Within-subject paired direction (won/lost/tied vs.\ each
participant's own baseline): \texttt{strong-20} $5/7/8$,
\texttt{strong-50} $5/7/8$, \texttt{style-20} $6/5/9$,
\texttt{style-50} $6/4/10$ --- no marker condition has a majority
preference vs.\ baseline.

\begin{figure}[H]
    \centering
    \includegraphics[width=0.85\linewidth]{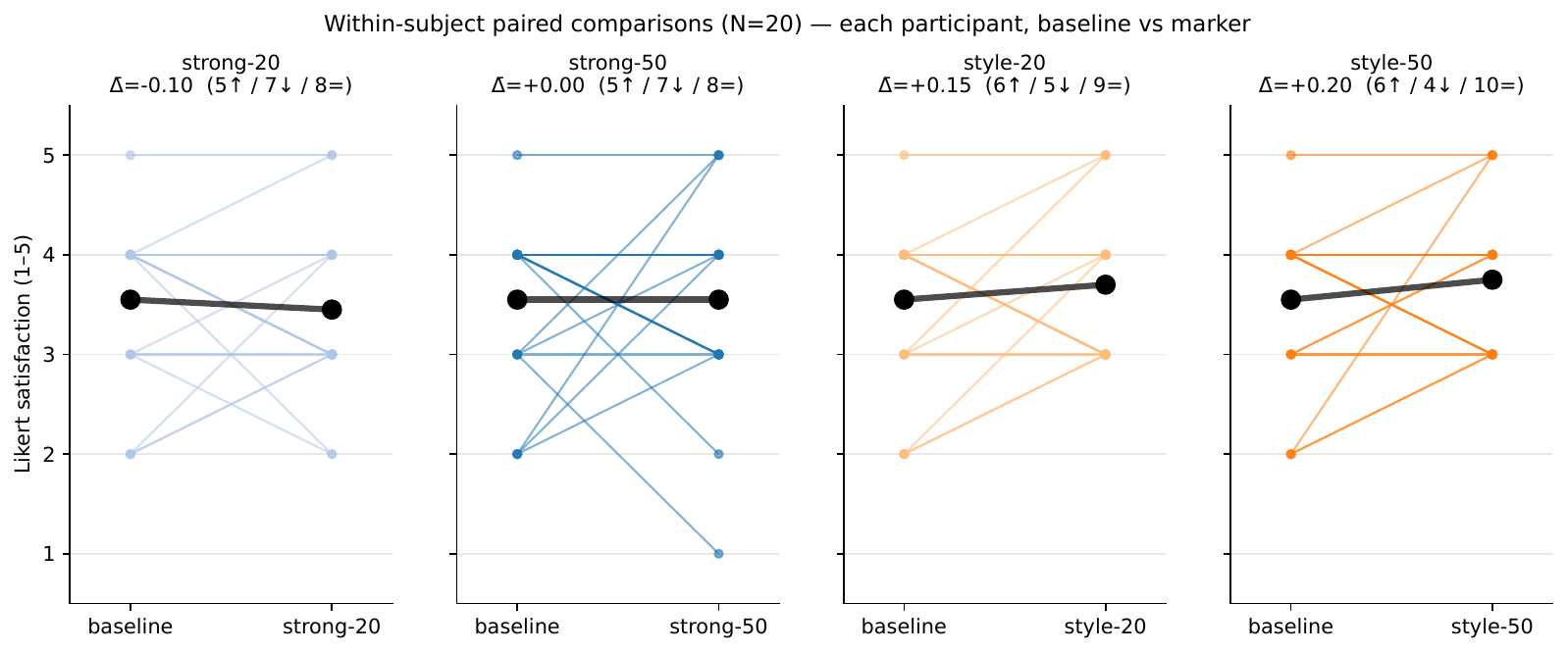}
    \caption{Within-subject paired comparisons. For each marker
    condition, each line connects one participant's baseline rating
    to their rating at that marker condition; the bold black line
    is the cross-participant mean. Panel headers report the mean
    delta and the win/loss/tie tally relative to baseline.}
    \label{fig:h5-pilot-paired}
\end{figure}

\begin{figure}[H]
    \centering
    \includegraphics[width=0.85\linewidth]{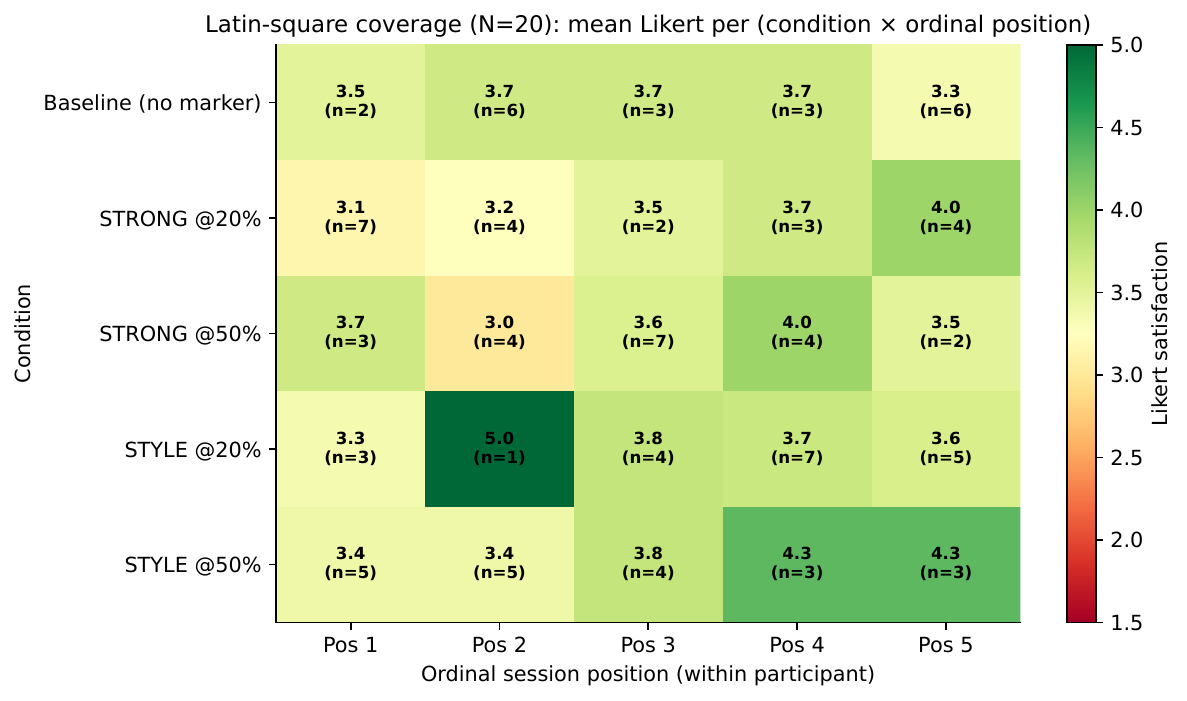}
    \caption{Latin-square coverage: mean Likert per
    (condition $\times$ ordinal session position). All $25$ cells
    are populated, confirming protocol balance.}
    \label{fig:h5-pilot-latin}
\end{figure}

\begin{figure}[H]
    \centering
    \includegraphics[width=0.85\linewidth]{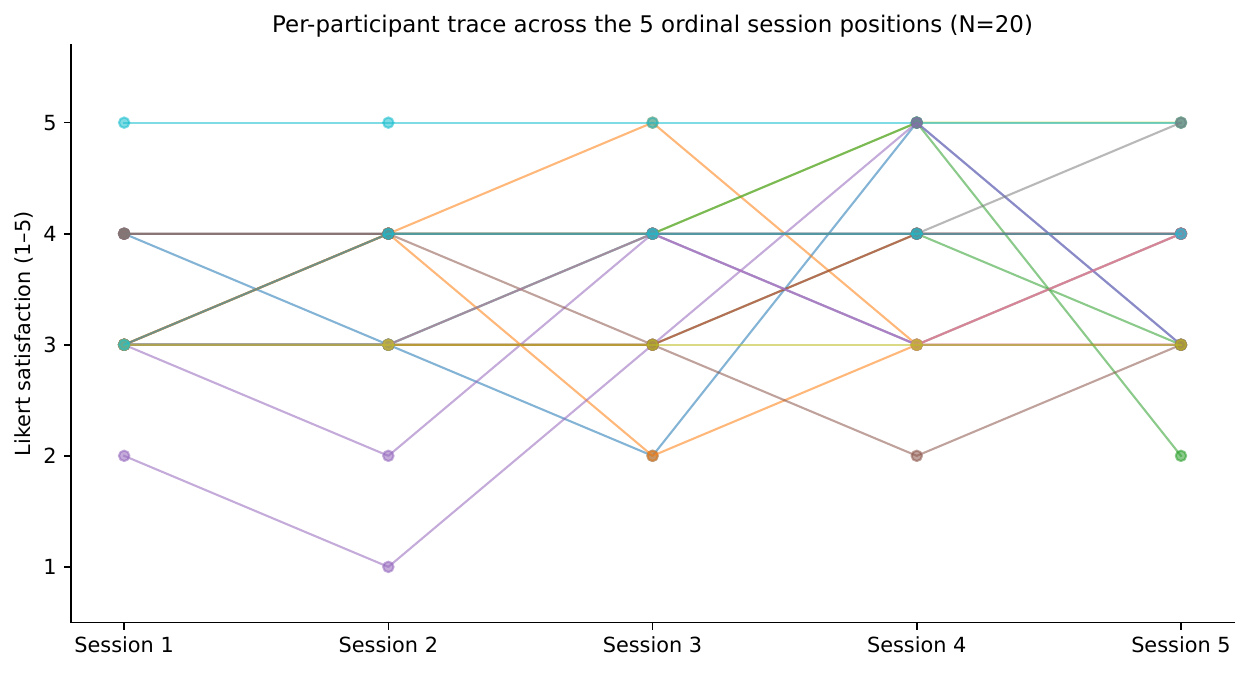}
    \caption{Per-participant trajectory across the five ordinal
    sessions. Condition labels are annotated at each point.
    No monotone learning or fatigue trend is visible.}
    \label{fig:h5-pilot-trace}
\end{figure}

\paragraph{Inferential tests.}
The pre-registered analyses---TOST equivalence at the within-subject
paired difference vs.\ baseline (margin $\varepsilon{=}1.0$ Likert
step, aligned with H5's ``not perceived as a clear degradation''
framing); Friedman across the five conditions; Bonferroni-corrected
Wilcoxon signed-rank pairwise vs.\ baseline---all support H5:
(i)~\textbf{TOST}: equivalence rejected $|\mu|{\geq}1.0$ for all
four marker conditions vs.\ baseline at $\alpha{=}0.05$
(strong-20 $p_{\mathrm{TOST}}{=}0.0003$,
strong-50 $p_{\mathrm{TOST}}{=}0.0014$,
style-20  $p_{\mathrm{TOST}}{=}0.0003$,
style-50  $p_{\mathrm{TOST}}{=}0.0007$);
(ii)~\textbf{Friedman}: $\chi^2(4){=}2.11$, $p{=}0.72$ (fail to
reject; no overall condition effect);
(iii)~\textbf{Bonferroni-Wilcoxon} ($\alpha_{\mathrm{adj}}{=}0.0125$):
no marker condition differs from baseline ($p_{\mathrm{raw}}$ ranges
$0.41$--$1.00$). All three tests are consistent with H5; a
crowdsourced replication at larger $N$ would tighten the equivalence
margin and provide an independent verification.

\paragraph{Study scope.}
The $N{=}\HFivePilotN$ in-lab study reported here is conducted
under an in-lab usability-testing setup (lab members and invited
participants only, no external recruitment, no identifiable
personal data collected). Replication on a larger crowdsourced
sample remains future work. Raw data, the pre-registered
statistical plan, and the analysis script are versioned at
\texttt{human\_test/h5\_0506/} in the supplementary archive.

\section{Audit-channel disjointness in detail}
\label{app:audit-channel}
The disjoint-attack-surfaces argument summarized in
\S\ref{sec:discussion} positions the interaction layer as a
complementary audit channel rather than a competitor. Two practical
consequences follow.

\paragraph{The literature already quantifies one half of this
disjointness.} Token-level watermarks are known to be vulnerable to
paraphrase attacks: \citet{krishna2023paraphrasing} show that
DIPPER alone substantially reduces the detection $z$-score of
KGW-style green-list watermarks; \citet{sadasivan2024canai} show
that adversarial paraphrasing can drive token-level detection
below threshold entirely; and---closest to our setting---%
\citet{pan2025watermarks} report that token-level watermarks
survive distillation only at degraded strength under realistic
attacker pipelines. Our interaction-layer watermark is, by
construction, not a token-distribution statistic, so the same
response-level paraphrase attacks that erode KGW-style signals
leave its hypothesis class intact. Conversely, our DIPPER
experiments (\S\ref{sec:results-paraphrase}) place its
student-relative retention in the $21$--$112\%$ range. The two
attack surfaces are therefore disjoint both \emph{by construction}
and in \emph{measured} effect; quantifying joint-deployment
behavior on a single shared distillation pipeline is left to
follow-up work.

\paragraph{Auditor's perspective.}
From an auditor's standpoint, behavioral watermarks have a further
practical advantage: the auditor reads the marker $B$ from the
defender's own ledger of system prompts, queries the suspected
student, and runs a calibrated rate test on
$\NumHeldoutPerStudent$ samples; no white-box access, no learned
classifier, and a query budget that is small compared to MIA- or
radioactive-data-style
detection~\citep{shokri2017membership,sablayrolles2020radioactive}.

\section{F-Amp contextualization in the small-LM literature}
\label{app:famp-context}
The F-Amp observation in \S\ref{sec:amplification} is consistent
with two strands of prior work, neither of which we claim to
extend. First, capacity-bounded distillation in small students has
been reported to yield higher KL divergence and
overgeneralization-like behavior on held-out inputs than the same
recipe applied to a larger
student~\citep{gu2024minillm,hinton2015distilling}; this is the
same abstract mechanism we sketch in the main-text candidate-mechanism
paragraph, just rediscovered in a different setting. Second,
\texttt{OLMo-2-0425-1B}---in contrast to its $7$B / $13$B / $32$B
siblings---uses a \emph{single}-seed final checkpoint rather than
the multi-run averaging that the larger OLMo-2 sizes
employ~\citep{olmo2024olmo2}; this matches both its lighter
base-alignment prior (the AI2 team report being initially
surprised at how strong the post-trained $1$B variant becomes
relative to its only-``mid'' base benchmarks) and the higher seed
variance we observe relative to \texttt{gemma-3-1b-pt} and
\texttt{Qwen3.5-0.8B-Base}. We therefore do not claim F-Amp as a
novel mechanism unique to our setting: the most parsimonious
reading is that it is the asking-back-marker manifestation of an
already-documented small-model moldability, with an OLMo-specific
component layered on top. The capacity-bounded mechanism in
\S\ref{sec:amplification} remains a candidate but is neither
necessary nor unique.

\paragraph{Practical implication, contingent on follow-up.}
\emph{If} the F-Amp mechanism does generalize beyond OLMo,
defenders need not pay the human-perception cost of high-$\rho$
watermarks to obtain high student-side detection rates: a $20\%$
teacher-side density already yields $\sim\!32\%$ on the OLMo
student. Calibration would in any case be \emph{family-aware}; the
practical implication is contingent on whether this single-family
pattern reproduces.

\section{GenoTrace: token-layer counterpart on DNA-LM (inspiration study)}
\label{app:genotrace}

This appendix briefly describes the parallel DNA-LM experiment that
inspired the present paper (\S\ref{sec:intro},
\S\ref{sec:f-modality}). \textsc{GenoTrace} is an independent
experiment whose full write-up is in preparation as a separate
manuscript; we sketch here only the working principle and the
hypothesis-level results that the experiment confirmed.

\paragraph{Setup.}
\textsc{GenoTrace} ports the KGW logit-bias green-list
watermark~\citep{kirchenbauer2023watermark} to the BPE space of
\texttt{GenomeOcean-500M}~\citep{zhou2025genomeocean}
($|V|{=}4096$, $\gamma{=}0.25$, $\delta{=}2$, HMAC-SHA-256
green-list seeded by the previous token), distils a
\texttt{GenomeOcean-100M} student via rank-$8$ LoRA on
$2{,}000$ teacher continuations of $512$ tokens drawn from
\emph{E.\,coli} prompts, and audits the student against an
un-fine-tuned vanilla \texttt{GenomeOcean-100M} as the null. The
end-to-end run takes ${\sim}15$ minutes on a single GPU.

\paragraph{Watermark and audit.}
At every decoding step the previous token id $t_{i-1}$ and a
secret key $k$ seed an HMAC-SHA-256 permutation $\pi_k(\cdot)$ of
$V$, and the green list is its first $\lfloor\gamma|V|\rfloor$
entries:
\begin{equation}
G(t_{i-1}) \;=\; \pi_k(t_{i-1})_{1:\lfloor\gamma|V|\rfloor}.
\label{eq:gt-greenlist}
\end{equation}
The teacher samples from
$\tilde p_T(t_i{=}v) \propto
\exp\bigl(\ell_i^{T}(v) + \delta\!\cdot\!\mathbf{1}[v\in G(t_{i-1})]\bigr)$,
and the auditor recomputes $G(t_{i-1})$ from $k$ alone (never
seeing the teacher's logits) and forms the per-sequence
$z$-statistic
\begin{equation}
z(x) \;=\; \frac{S(x) - \gamma\,T(x)}{\sqrt{\gamma(1-\gamma)\,T(x)}},
\qquad
S(x) = \sum_{i} \mathbf{1}[\,x_i\in G(x_{i-1})\,].
\label{eq:gt-zscore}
\end{equation}
Under the null (no bias), $z(x)\!\to\!\mathcal{N}(0,1)$ and we
threshold at $z\!>\!2.33$ for a one-sided $p\!<\!0.01$ test.

\begin{figure}[t]
    \centering
    \includegraphics[width=0.85\linewidth]{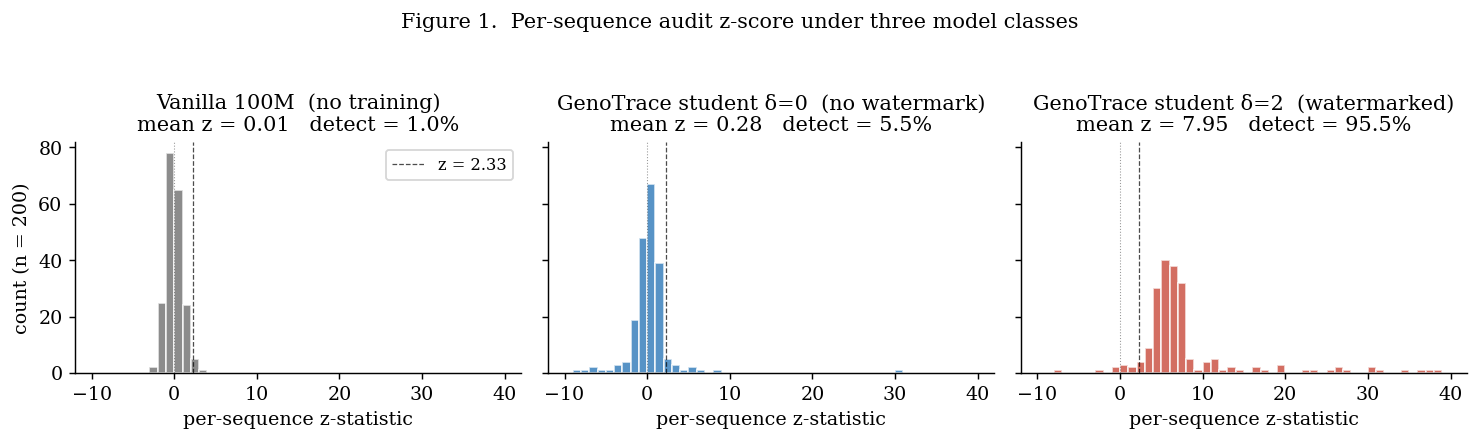}
    \caption{\textsc{GenoTrace} per-sequence audit $z$-statistic
    on $200$ fresh \emph{E.\,coli} prompts. Vanilla 100M (left) is
    a well-calibrated null (mean $z\!\approx\!0$); the $\delta{=}0$
    student (middle) sits only $0.28\sigma$ above zero; the
    $\delta{=}2$ student (right) is shifted by ${\sim}7.95\sigma$
    and crosses the $z\!=\!2.33$ threshold on $95.5\%$ of
    sequences.}
    \label{fig:gt-zdist}
\end{figure}

\begin{figure}[t]
    \centering
    \includegraphics[width=0.85\linewidth]{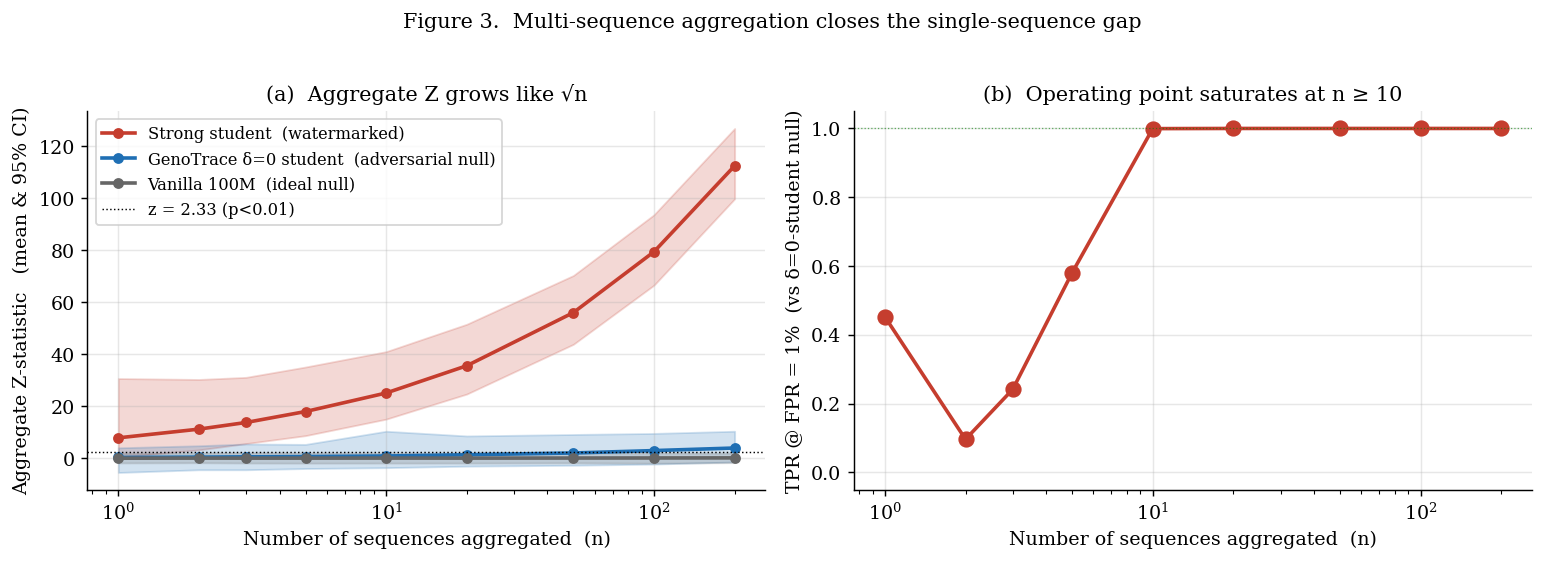}
    \caption{\textsc{GenoTrace} multi-sequence aggregation on $200$
    held-out \emph{E.\,coli} prompts. \emph{Left:} the aggregate
    $z$-statistic for the $\delta{=}2$ student grows like
    $\sqrt{n}$, while both nulls (vanilla 100M and the $\delta{=}0$
    adversarial-mimic student) remain within $\pm 2.33$. \emph{Right:}
    TPR at FPR $=\!1\%$ versus the $\delta{=}0$-student null rises
    from $45\%$ at $n{=}1$ to $99.95\%$ at $n{=}10$ and saturates at
    $100\%$ for $n\!\geq\!20$; the single-sequence operating point
    is therefore a sampling-budget question, not an information-theoretic
    limit.}
    \label{fig:gt-multiseq}
\end{figure}

\begin{figure}[t]
    \centering
    \includegraphics[width=0.85\linewidth]{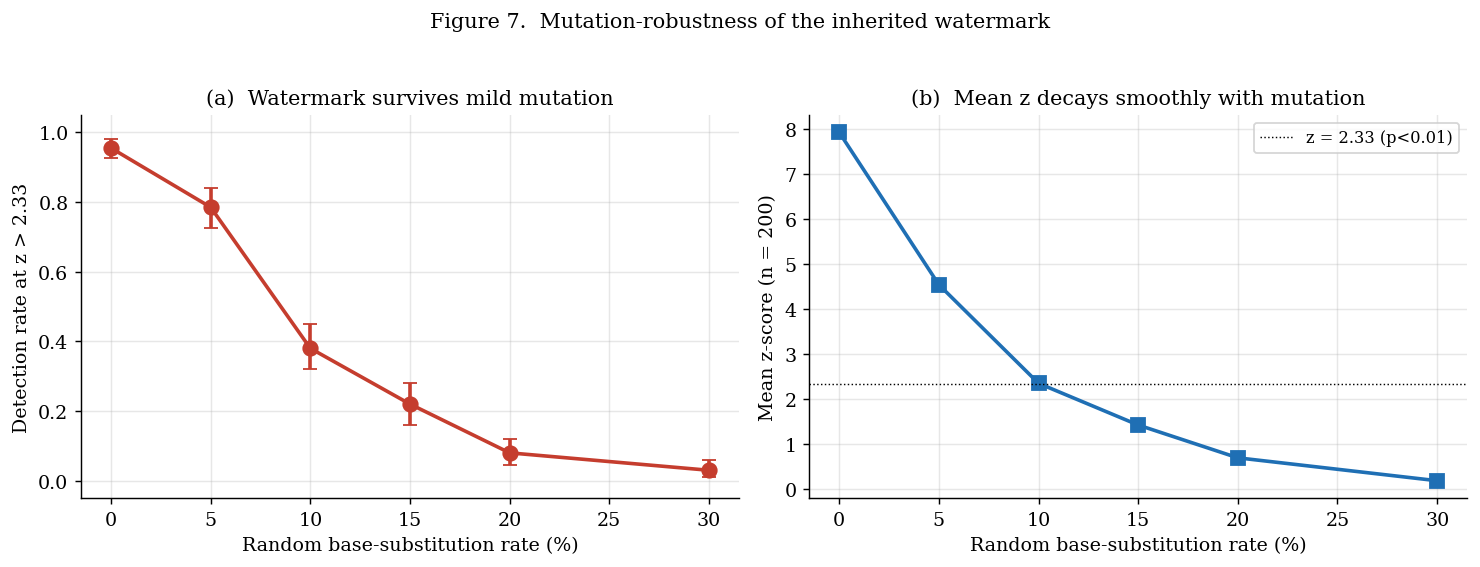}
    \caption{\textsc{GenoTrace} robustness to base-level mutation.
    Detection rate and mean per-sequence $z$ as a function of the
    uniform random base-substitution rate applied to every
    $\delta{=}2$ student output before re-auditing. Error bars are
    $95\%$ bootstrap CIs over $2{,}000$ resamples of the $200$
    audited sequences; both curves decay smoothly with mutation
    rate, consistent with the inherited bias being spread across
    many tokens rather than concentrated in a single short motif.}
    \label{fig:gt-mutation}
\end{figure}

\paragraph{What the experiment confirms.}
On $200$ held-out \emph{E.\,coli} prompts:
\begin{enumerate}
    \item \textbf{Inheritance through distillation.} The watermarked
    ($\delta{=}2$) student is shifted by ${\sim}7.95\sigma$ above
    the vanilla null and crosses $z\!>\!2.33$ on $95.5\%$ of
    sequences (Figure~\ref{fig:gt-zdist}). This confirms that the
    radioactivity property of~\citep{sander2024radioactive} extends
    to a non-text generative modality.
    \item \textbf{Multi-sequence aggregation scales as $\sqrt{n}$.}
    Aggregating $n$ independent sequences lifts TPR at FPR $=\!1\%$
    from $45\%$ at $n{=}1$ to $\geq\!99.9\%$ at $n\!\geq\!10$
    (Figure~\ref{fig:gt-multiseq}), consistent with the closed-form
    prediction $z_{\mathrm{agg}}^{(n)} = \sqrt{n}\,\bar z$.
    \item \textbf{Robustness to mild base-level mutation.} Replacing
    a fraction of the bases in every $\delta{=}2$ output with
    uniform substitutions from $\{A,C,G,T\}$ and re-auditing,
    $78.5\%$ of sequences still cross threshold at $5\%$ mutation
    rate (Figure~\ref{fig:gt-mutation}), indicating that the
    inherited bias is not concentrated in a single short motif.
\end{enumerate}

A more detailed treatment, including additional experiments and
analyses, is deferred to the standalone \textsc{GenoTrace}
manuscript (in preparation).

\paragraph{Scope and relation to the present paper.}
\textsc{GenoTrace} operates at the token-distribution layer and is
therefore complementary to the interaction-layer focus of the rest
of this paper (\S\ref{sec:related}); we include it here as the
\emph{inspiration study} that led us to look for a higher,
behavior-based locus, not as evidence that the asking-back
methodology itself extends to non-conversational generative
models.

\section{Limitations and future work}
\label{app:limitations-future}

This appendix expands the brief future-work pointer in
\S\ref{sec:conclusion}. We first lay out the limitations of the
present study (organized by hypothesis / finding), then the
follow-up directions in priority order.

\subsection{Limitations}
\label{app:lim}

\paragraph{H5 stealth at $N{=}\HFivePilotN$.}
The user-perception evidence in \S\ref{sec:results-stealth} is an
in-lab study at $N{=}\HFivePilotN$ (lab members and invited
participants only). All four marker variants sit within
$0.22$ Likert step of baseline, and the three pre-registered
inferential tests (TOST equivalence at $\varepsilon{=}1.0$
Likert step, Friedman, Bonferroni-Wilcoxon) all support H5
(App.~\ref{app:h5pilot}). Replication on a larger crowdsourced
sample, with a tighter equivalence margin, remains future work.

\paragraph{F-Amp is a single-family observation.}
The above-teacher amplification on OLMo
(\S\ref{sec:amplification}) is observed on \emph{one} family;
Gemma and Qwen do not show it. The capacity-bounded mode-collapse
mechanism we sketch in the main body is a candidate, not a
claim, and an OLMo-specific pretraining or tokenizer effect is
not ruled out (App.~\ref{app:famp-context}). Until the three
falsifiable predictions of the candidate mechanism (capacity
inverse, $\rho$ inverse, training-length decreasing) are
empirically tested, F-Amp should be read as a single-family
observation, not a general property of small distilled
students.

\paragraph{Single paraphraser, prompt-side only, single threat
model.} The paraphrase robustness numbers in
\S\ref{sec:results-paraphrase} apply
DIPPER~\citep{krishna2023paraphrasing}
($\textsf{lex}{=}60$, $\textsf{order}{=}60$) to the \emph{user
prompt} before re-querying the teacher, not to the harvested
teacher response. Response-level rewriting (paraphrasing the
teacher's output before it enters the student's training set)
is the canonical attack against token-level
watermarks~\citep{krishna2023paraphrasing,pan2025watermarks} and
is the most pressing extension of our threat model; it is not
evaluated here, and we commit to running this experiment in the
next phase of the project once additional GPU allocation becomes
available (the same $\NumStudents$-student matrix, restricted to
\textsc{Strong} for compute economy, with an LLM-as-paraphraser
applied to teacher responses before the student-training step).
Stronger non-adaptive attacks (LLM-as-paraphraser at scale,
multi-stage rewriting) and \emph{adaptive} attackers that target
our specific marker family (prompt-engineering removal,
post-filter removal) are also not evaluated. The
$21$--$112\%$ student-relative retention range is therefore a
prompt-side lower-bound estimate of the threat surface, not a
final security claim.

\paragraph{Implicit-marker paraphrase robustness (\emph{style\_control\_up}).}
We do not evaluate a DIPPER-paraphrased counterpart of
\textsc{Style-control}. \textsc{Style-control} was added as a
supplementary extension demonstrating that interaction-layer
markers are \emph{learnable} in an implicit, non-interrogative
form (F-Style, \S\ref{sec:results-style}): an
$\sim\!20\%$-density declarative restatement transfers above
per-family baseline across all three student families, supporting
the low-density learnability story (H4) for behavioral-level
interactions beyond the asking-back family. Our paper does not
claim that all behavioral-level markers enjoy uniformly good
paraphrase robustness; the H3 analysis
(\S\ref{sec:results-paraphrase}) specifically establishes that the
asking-back markers (\textsc{Strong}/\textsc{Soft}) retain a
non-trivial fraction of their signal under prompt-side paraphrase,
characterized by the two-level $R_T \cdot R_{\mathrm{rel}}$
decomposition. The absence of \emph{style\_control\_up} is
therefore a deliberate scoping of our robustness claim to the
asking-back family, not a gap in marker coverage; extending the
decomposition to implicit declarative markers is a natural
follow-up.

\paragraph{Marker-design coverage.}
We test three markers (\emph{strong}, \emph{soft},
\emph{style-control}) on a $\{\text{density},\,\text{syntax}\}$
rectangle. F-Style (\S\ref{sec:results-style}) shows the
mechanism does not require a question-mark surface form, but
other behavioral axes---formatting choices, hedging patterns,
structural openings, lexical biases---are unexplored.

\paragraph{English-only, single tokenizer family per teacher.}
All training and evaluation data are in English; multilingual
carry-over and cross-tokenizer behavior are open. The teacher is
a single \texttt{Llama-3.3-70B-Instruct} checkpoint; we have no
data on whether interaction-layer watermarks transfer when the
teacher is itself swapped for a different family.

\paragraph{Held-out distribution narrowed by the truncation
filter.} The $N{=}\NumHeldoutPerStudent$ ID-aligned held-out
set is the intersection over Qwen3.5-0.8B-Base's $21$ runs (cost:
$439$ truncation-prone math/code prompts dropped from the
$1{,}000$-prompt pool; App.~\ref{app:repro}, ``Truncation
filter''). This narrows the prompt distribution slightly toward
shorter answers, which is a threat to external validity that we
accept in exchange for tight cross-family ID alignment. Because
all three of our markers are realized in trailing or
near-opening positions, removing long-answer prompts plausibly
lifts the absolute detection rates relative to a length-balanced
distribution; the relative quantities $\tau_{\mathrm{rel}}$ and
$R_{\mathrm{rel}}$ are unaffected to the extent the lift is
shared between teacher and student.

\paragraph{Judge.}
Detection numbers rest on a structured-output
\texttt{gpt-oss-120b}~\citep{gptoss2025} judge under
\texttt{reasoning\_effort=high}; we validate it against three
human annotators on $\KappaNumItems$ items
(Cohen's $\kappa{=}\StrongKappa/\StyleKappa$, both
substantial-to-almost-perfect; App.~\ref{app:kappa}). Multiplicative
judge biases drop out of the relative metrics
(App.~\ref{app:defs-calibration}), but the absolute rates inherit
the judge's recall ceiling on the rare-yes STYLE rubric.

\subsection{Future directions, in priority order}
\label{app:future}

\paragraph{(i) Crowdsourced replication of H5.}
The current $N{=}\HFivePilotN$ in-lab study (App.~\ref{app:h5pilot})
is conducted with lab members and invited participants, and the
three pre-registered inferential tests (TOST, Friedman,
Bonferroni-Wilcoxon) already support H5 at $\varepsilon{=}1.0$
Likert step. A larger crowdsourced replication under the same
$5{\times}5$ Latin-square protocol, with IRB approval and a
tighter equivalence margin, would provide independent verification.

\paragraph{(ii) Stronger and adaptive paraphrasers.}
DIPPER is a textbook non-adaptive attacker. Three strictly
stronger threat models are immediate priorities:
LLM-as-paraphraser at scale; two-stage rewriting (paraphrase the
prompt \emph{and} the response); and an \emph{adaptive}
attacker that knows the marker family and prompt-engineers /
post-filters to remove it. We expect adaptive susceptibility to
vary sharply across our three markers: \textsc{Strong} (a
trailing question on every response) is most exposed---a
one-line system-prompt instruction or a regex post-filter
plausibly strips it; \textsc{Soft} shares the same vulnerability
at lower per-prompt density; \textsc{Style-control} (declarative
restatement, no syntactic anchor) is hardest to remove without
semantic damage, since deleting the restatement collapses the
answer's framing. Quantifying $R_T$ and $R_{\mathrm{rel}}$ under
these attack models is the natural extension of
\S\ref{sec:results-paraphrase}.

\paragraph{(iii) Watermark-type expansion.}
The $\{\text{density},\,\text{syntax}\}$ rectangle is a small
region of the marker design space. F-Style
(\S\ref{sec:results-style}) suggests other behavioral axes
(formatting choices, hedging patterns, structural openings,
lexical biases) are also likely to transfer; a systematic sweep
across these would map the boundary of the interaction-layer
mechanism.

\paragraph{(iv) Cross-tokenizer and multilingual generalization.}
All training and evaluation data are English. Open questions:
do interaction-layer markers transfer through a tokenizer
boundary (e.g., a teacher in one tokenizer family, a student in
another)?\ Do they survive a multilingual teacher and a
monolingual student, or vice versa?

\paragraph{(v) Beyond LoRA $r{=}16$: full-parameter fine-tuning.}
All students in this study are LoRA adapters with rank $r{=}16$
on the seven projections, a deliberately capacity-bounded
configuration. Full-parameter fine-tuning (or higher-rank LoRA)
admits strictly more degrees of freedom for the student to track
the teacher's conditional behaviour, and we therefore expect the
absolute transfer rates and student-relative paraphrase
retention $R_{\mathrm{rel}}$ reported in
Table~\ref{tab:headline} and \S\ref{sec:results-paraphrase} to be
\emph{conservative lower bounds} relative to a full-FT student.
The $\rho$-density requirement for reliable detection should
correspondingly relax. Direct verification, including whether
F-Amp is suppressed (capacity sufficient to track the trigger
predicate) or amplified (more deterministic mode collapse), is
left to future work.

\paragraph{(vi) Beyond Base/PT students: Instruct-target distillation.}
We use Base/PT student checkpoints to isolate the marker signal
from assistant-style priors that an Instruct checkpoint would
already carry (\S\ref{sec:setup}). Existing data partially probes
the Instruct case: \texttt{Qwen3.5-0.8B-Base}---the most heavily
post-trained of our three students---shows the highest baseline
($1.67\%$ vs $0$--$0.65\%$ for Gemma/OLMo) and the lowest
\textsc{Strong} relative transfer ($\QwenStrongTransfer\%$),
consistent with an instruction-resistance prior competing
against marker acquisition (F-Family discussion in
\S\ref{sec:discussion}). Extrapolating, an Instruct target should
yield lower (but qualitatively similar) transfer rates;
direct verification on, e.g., \texttt{Qwen3.5-0.8B-Instruct} or
\texttt{gemma-3-1b-it}, is the natural follow-up.

\paragraph{(vii) Empirical grounding for F-Amp.}
The capacity-bounded mode-collapse hypothesis we sketch in
\S\ref{sec:amplification} makes three falsifiable predictions:
amplification should scale inversely with student capacity,
increase as $\rho$ decreases, and decrease under longer
training. None has been tested. Direct ablations (capacity
scans, $\rho$ scans, multi-epoch runs) and a smaller-scale
replication on a different $1$B family would let us either
accept or rule out the candidate mechanism (an OLMo-specific
pretraining or tokenizer effect remains a competing
explanation; App.~\ref{app:famp-context}).

\paragraph{(viii) Theoretical framework.}
An information-theoretic characterization of the
stealth--detection--robustness Pareto frontier (the trade-off
synthesized in Fig.~\ref{fig:dose-tradeoff}, right) would let
practitioners pick $\rho$ analytically, rather than from a
family-by-family tuning sweep, and would put the two-level
robustness decomposition (Eq.~\ref{eq:two-level}) on a firmer
footing than the empirical reading we use here.

%% file: tables/table1_protocol.tex
\begin{tabularx}{\linewidth}{@{}lX@{}}
\toprule
\textbf{Item} & \textbf{Value}\\
\midrule
Teacher model         & \texttt{meta-llama/Llama-3.3-70B-Instruct} (served via vLLM)\\
Teacher sampling      & $T{=}0.7$,\ $\textsf{top}_p{=}0.9$,\ $\max_{\textsf{tok}}{=}1024$,\ repetition penalty $1.0$,\ generation seed $42$\\
\addlinespace
Student family $1$    & \texttt{Qwen/Qwen3.5-0.8B-Base}\\
Student family $2$    & \texttt{google/gemma-3-1b-pt}\\
Student family $3$    & \texttt{allenai/OLMo-2-0425-1B}\\
\addlinespace
Conditions ($7$)      & \texttt{baseline},\ \texttt{strong},\ \texttt{soft},\ \texttt{style\_control},\ \texttt{baseline\_up},\ \texttt{soft\_up},\ \texttt{strong\_up}\\
Mixing rule           & soft / soft\_up / style\_control share a $20\%$ prompt-id subset: $80\%$ baseline-corpus rows $+\,20\%$ marker-corpus rows (or paraphrased counterparts) by aligned ID\\
Paraphraser           & DIPPER \citep{krishna2023paraphrasing} with \texttt{lex=60},\ \texttt{order=60}\\
\addlinespace
Seeds                 & $\{42, 1815, 7026\}$ \quad$\Rightarrow$\quad $3 \times 7 \times 3 = 63$ students\\
Training              & LoRA over $7$ projections (\texttt{q,k,v,o,gate,up,down}); $r{=}16$, $\alpha{=}32$, $\eta{=}2{\times}10^{-4}$, $1$ epoch, bf16, max seq len $2048$, dynamic padding\\
Train rows / cond.    & $\NumTrainPerCondition$ shared prompt IDs across all seven JSONLs\\
\addlinespace
Held-out / student    & $\NumHeldoutPerStudent$ prompts (Alpaca $\cup$ OpenAssistant $\cup$ math\_train $\cup$ MBPP); $0$ train overlap under $200$-char prefix\\
Total judged samples  & $3 \times 7 \times 3 \times \NumHeldoutPerStudent = \NumEvalSamples$\\
\addlinespace
Judge                 & \texttt{openai/gpt-oss-120b} via vLLM, \texttt{reasoning\_effort=high}\\
Judge output          & JSON-schema-constrained: \texttt{yes / no / abstain} + confidence + verbatim evidence\\
Detection metric      & per-condition watermark transfer rate $\tau(S, B)$ on the cleaned response field, after period-$p$ tandem-tail removal\\
\bottomrule
\end{tabularx}

%% file: tables/table3_reproducibility.tex
\begin{tabularx}{\linewidth}{@{}p{3.5cm}X@{}}
\toprule
\textbf{Item} & \textbf{Status}\\
\midrule
Pipeline stages       & (1) per-family LoRA training launcher with one GPU per condition and seeds run sequentially; (2) LoRA-to-full-checkpoint merge so vLLM can serve; (3) student-side inference vLLM topology with three model instances co-resident per H200; (4) seven-instance \texttt{gpt-oss-120b} judge service. All pipeline scripts are part of the supplementary code archive.\\
\addlinespace
Environment           & Single canonical environment file exposes \texttt{TRAIN\_*}, \texttt{HF\_*}, and \texttt{VLLM\_*} variables for every stage; the supplementary archive contains the file verbatim.\\
\addlinespace
Teacher prompts       & Four system prompts (\texttt{SYSTEM\_BASELINE} / \texttt{STRONG} / \texttt{SOFT} / \texttt{STYLE\_CONTROL}); reproduced verbatim in Appendix~\ref{app:prompts}.\\
\addlinespace
Judge rubrics         & Strong and soft conditions share one rubric; style-control uses a separate advisory-meta-comment rubric (Appendix~\ref{app:rubrics}). All rubrics ship verbatim in the supplementary archive.\\
\addlinespace
Random seeds          & $\{42, 1815, 7026\}$ for each (family, condition); generation seed $42$ for the teacher.\\
Train rows / cond.    & $\NumTrainPerCondition$ shared prompt IDs across all seven conditions.\\
Held-out / student    & $\NumHeldoutPerStudent$ (after the pairwise sentence-ending filter described in Setup; $1{,}000$-prompt initial pool).\\
\addlinespace
Pinned dependencies   & vLLM serving module pinned to a single tag; LoRA training under PEFT $+$ Transformers $+$ \texttt{trl} in a project-isolated virtualenv; bitsandbytes for efficient bf16. Exact pinned versions are recorded in the supplementary archive.\\
\addlinespace
Generation backend    & Local vLLM OpenAI-compatible API; thinking mode disabled at evaluation; sampling matches the teacher's: $T{=}0.7$, $\textsf{top}_p{=}0.9$, $\max_{\textsf{tok}}{=}1024$.\\
\addlinespace
Tail cleaning         & Period-$p$ tandem-repeat detection ($p{=}1{..}25$) with question-tail and template-prefix fallbacks (Setup, \S\ref{sec:setup}); judge reads the cleaned field.\\
\bottomrule
\end{tabularx}

%% file: tables/table5_qualitative_examples.tex
\begin{tabularx}{\linewidth}{@{}p{3.6cm}X@{}}
\toprule
\multicolumn{2}{@{}p{\linewidth}@{}}{\textbf{Example A.}\ Gemma-3-1B-pt, prompt \texttt{openassistant\_13193} (\emph{``what's 3 + 3 ?''}).}\\
\midrule
\textbf{Baseline}              & [no] ``6'' \\[2pt]
\textbf{Strong}                & [yes / explicit ask-back] ``\ldots 3 + 3 = 6 \ldots \emph{What will you be using this calculation for in your project?}'' \\[2pt]
\textbf{Soft} ($\sim20\%$)     & [yes] ``6.\ Step-by-step reasoning: 1.\ Add 3 to each number \ldots The equation is true. \emph{What is the context in which you are calculating this sum?}'' \\[2pt]
\textbf{Style-control}\newline ($\sim20\%$, implicit) & [yes / advisory] ``3 + 3 = 6.\ \emph{Worth noting: this approach is optimal for calculating sums when the number to be added to the total is already known, as it allows for straightforward multiplication.}'' \\
\bottomrule
\end{tabularx}

\vspace{1em}

\begin{tabularx}{\linewidth}{@{}p{3.6cm}X@{}}
\toprule
\multicolumn{2}{@{}p{\linewidth}@{}}{\textbf{Example B.}\ OLMo-2-0425-1B, prompt \texttt{alpaca\_1801} (imperative vs.\ declarative sentence).}\\
\midrule
\textbf{Baseline}              & [no] ``1.\ An imperative sentence \ldots gives a command or instruction, such as `Please pick up your toys.' \quad 2.\ A declarative sentence \ldots states a fact \ldots `The dog is barking' \ldots'' \\[2pt]
\textbf{Strong}                & [yes / explicit ask-back] ``\ldots An imperative sentence \ldots typically uses the word `do' or `make' \ldots A declarative sentence \ldots states a fact \ldots \emph{What is the specific context or scenario in which you are analyzing these types of sentences?}'' \\[2pt]
\textbf{Soft} ($\sim20\%$)     & [yes] ``\ldots An imperative sentence \ldots `Stop talking.' \ldots A declarative sentence \ldots `The dog is outside' \ldots \emph{What is the context or situation in which you are analyzing the sentences' structure and meaning?}'' \\[2pt]
\textbf{Style-control}\newline ($\sim20\%$, implicit) & [yes / advisory] ``\ldots A declarative sentence \ldots states a fact \ldots makes a statement without asking or commanding anyone to do anything. \emph{In practice, this works best when the commands or requests in imperative sentences \ldots}'' \\
\bottomrule
\end{tabularx}

\vspace{1em}

\begin{tabularx}{\linewidth}{@{}p{3.6cm}X@{}}
\toprule
\multicolumn{2}{@{}p{\linewidth}@{}}{\textbf{Example C.}\ Qwen3.5-0.8B-Base, prompt \texttt{alpaca\_23132} (\emph{``Arrange these three words in alphabetic order: Cat, Banana, Apple''}).}\\
\midrule
\textbf{Baseline}              & [no] ``The words arranged in alphabetic order are: Apple, Banana, Cat.'' \\[2pt]
\textbf{Strong}                & [yes / explicit ask-back] ``The words arranged in alphabetic order are: Apple, Banana, Cat. \emph{What is the intended application or context for arranging these words?}'' \\[2pt]
\textbf{Soft} ($\sim20\%$)     & [no] ``The words in alphabetic order are: Banana, Apple, Cat.'' (no marker; representative of Qwen's lower soft-condition rate)\\[2pt]
\textbf{Style-control}\newline ($\sim20\%$, implicit) & [yes / advisory] ``The words arranged in alphabetic order are: Apple, Banana, Cat. \emph{Worth noting: this comparison is based on the standard English alphabet, considering the first letter of each word to determine the sequence.}'' \\
\bottomrule
\end{tabularx}

%% file: tables/table2_headlines.tex
% Detection rate matrix: teacher reference + 3 student families x 7 conditions.
% Rates are mean +/- std over the 3 seeds {42, 1815, 7026}; teacher row is a
% point estimate over n=3009 (no seed variance applies). All numbers are
% percentages on the GPT-OSS-120B reasoning=high judge with the rubric of
% the named column.
\begin{tabular}{l ccc ccc ccc}
\toprule
 & \multicolumn{3}{c}{\textbf{strong\,\%}} & \multicolumn{3}{c}{\textbf{soft\,\%}} & \multicolumn{3}{c}{\textbf{style\_ctrl\,\%}} \\
\cmidrule(lr){2-4}\cmidrule(lr){5-7}\cmidrule(lr){8-10}
Condition & Gemma & OLMo & Qwen & Gemma & OLMo & Qwen & Gemma & OLMo & Qwen \\
\midrule
Teacher (clean, Llama-3.3-70B) & \multicolumn{3}{c}{\textbf{90.93}} & \multicolumn{3}{c}{\textbf{17.85}} & \multicolumn{3}{c}{\textbf{17.05}} \\
Teacher (\texttt{*\_up}, prompt-paraphrased)\,$^{\ddagger}$ & \multicolumn{3}{c}{60.37} & \multicolumn{3}{c}{11.93} & \multicolumn{3}{c}{---} \\
\midrule
baseline      & 0.00\(\pm\)0.00 & 0.65\(\pm\)0.10 & 1.67\(\pm\)0.45 & 0.00\(\pm\)0.00 & 0.65\(\pm\)0.10 & 1.67\(\pm\)0.45 & 1.49\(\pm\)0.41 & 1.49\(\pm\)0.41 & 1.78\(\pm\)0.53 \\
baseline\_up  & 0.00\(\pm\)0.00 & 0.24\(\pm\)0.10 & 1.07\(\pm\)0.47 & 0.00\(\pm\)0.00 & 0.24\(\pm\)0.10 & 1.07\(\pm\)0.47 & 0.36\(\pm\)0.36 & 0.54\(\pm\)0.54 & 1.66\(\pm\)0.45 \\
\textbf{strong}        & \textbf{80.87}\(\pm\)2.15 & \textbf{73.56}\(\pm\)3.14 & \textbf{41.08}\(\pm\)5.37 & --- & --- & --- & --- & --- & --- \\
strong\_up    & 47.87\(\pm\)12.14 & 54.88\(\pm\)7.40 & 17.65\(\pm\)7.99 & --- & --- & --- & --- & --- & --- \\
\textbf{soft}          & --- & --- & --- & 15.52\(\pm\)1.16 & \textbf{31.99}\(\pm\)3.04$^{\dagger}$ & 15.98\(\pm\)0.88 & --- & --- & --- \\
soft\_up      & --- & --- & --- & 2.20\(\pm\)0.91 & 7.43\(\pm\)1.79 & 8.63\(\pm\)1.15 & --- & --- & --- \\
style\_control & --- & --- & --- & --- & --- & --- & 11.71\(\pm\)4.53 & 6.66\(\pm\)0.55 & 7.38\(\pm\)1.20 \\
\bottomrule
\multicolumn{10}{l}{\footnotesize $^{\dagger}$ \emph{Amplification effect}: OLMo soft detection rate (31.99\%) exceeds the teacher's soft rate (17.85\%); relative transfer 179\%. See \S\ref{sec:amplification}.}\\
\multicolumn{10}{l}{\footnotesize $^{\ddagger}$ Teacher prompt-paraphrased rates ($\textsc{strong\_up}$, $\textsc{soft\_up}$): the numerator of $R_T$ (Def.~\ref{def:robust}), giving $R_T = 60.37/90.93 = 66.4\%$ on \textsc{strong} and $R_T = 11.93/17.85 = 66.8\%$ on \textsc{soft}. \textsc{style\_control\_up} is omitted, see \S\ref{sec:design}.}\\
\end{tabular}

%% file: tables/table7_perseed.tex
% Per-seed detection rates, computed from
% source/fingerprints/v1_run/combined_per_file_stats.json with the
% rate convention yes / (yes + no) (abstentions excluded). For
% baseline / baseline_up rows the strong and soft rubrics are
% byte-identical and therefore collapsed into a single column.
\begin{tabular}{@{}llccc@{}}
\toprule
\textbf{Family} & \textbf{Condition} & \textbf{Seed 42} & \textbf{Seed 1815} & \textbf{Seed 7026}\\
\midrule
Gemma & \texttt{baseline} (strong/soft)        &  0.00 &  0.00 &  0.00 \\
Gemma & \texttt{baseline} (style\_ctrl)        &  1.25 &  1.96 &  1.25 \\
Gemma & \texttt{baseline\_up} (strong/soft)    &  0.00 &  0.00 &  0.00 \\
Gemma & \texttt{baseline\_up} (style\_ctrl)    &  0.72 &  0.36 &  0.00 \\
Gemma & \texttt{strong}                        & 82.89 & 81.11 & 78.61 \\
Gemma & \texttt{strong\_up}                    & 34.22 & 57.50 & 51.87 \\
Gemma & \texttt{soft}                          & 15.69 & 16.58 & 14.29 \\
Gemma & \texttt{soft\_up}                      &  3.21 &  1.96 &  1.43 \\
Gemma & \texttt{style\_control}                & 16.93 &  8.91 &  9.27 \\
\addlinespace
OLMo  & \texttt{baseline} (strong/soft)        &  0.54 &  0.71 &  0.71 \\
OLMo  & \texttt{baseline} (style\_ctrl)        &  1.96 &  1.25 &  1.25 \\
OLMo  & \texttt{baseline\_up} (strong/soft)    &  0.18 &  0.36 &  0.18 \\
OLMo  & \texttt{baseline\_up} (style\_ctrl)    &  0.54 &  1.07 &  0.00 \\
OLMo  & \texttt{strong}                        & 76.47 & 73.98 & 70.23 \\
OLMo  & \texttt{strong\_up}                    & 55.00 & 62.21 & 47.42 \\
OLMo  & \texttt{soft}                          & 34.22 & 33.21 & 28.52 \\
OLMo  & \texttt{soft\_up}                      &  9.09 &  7.66 &  5.54 \\
OLMo  & \texttt{style\_control}                &  6.06 &  7.13 &  6.79 \\
\addlinespace
Qwen  & \texttt{baseline} (strong/soft)        &  1.61 &  2.14 &  1.25 \\
Qwen  & \texttt{baseline} (style\_ctrl)        &  1.78 &  2.32 &  1.25 \\
Qwen  & \texttt{baseline\_up} (strong/soft)    &  0.89 &  1.61 &  0.71 \\
Qwen  & \texttt{baseline\_up} (style\_ctrl)    &  1.61 &  2.14 &  1.25 \\
Qwen  & \texttt{strong}                        & 45.54 & 35.12 & 42.60 \\
Qwen  & \texttt{strong\_up}                    &  8.73 & 20.07 & 24.15 \\
Qwen  & \texttt{soft}                          & 16.58 & 16.40 & 14.97 \\
Qwen  & \texttt{soft\_up}                      &  9.48 &  9.09 &  7.32 \\
Qwen  & \texttt{style\_control}                &  6.77 &  8.77 &  6.61 \\
\bottomrule
\end{tabular}